\documentclass[twocolumn,english,showpacs,superscriptaddress,prl]{revtex4}
\usepackage{amsmath}
\usepackage{amssymb}
\usepackage{graphicx}
\usepackage{txfonts}
\usepackage{color}
\usepackage{graphics}
\usepackage{ams}
\usepackage{ulem}

\begin{document}
\draft
\title{\bf First-principles investigation of dynamical properties
of molecular devices under a steplike pulse}

\author{Yanxia Xing$^{1,2}$, Bin Wang$^{1}$ and Jian Wang$^{1,*}$}

\address{
$^1$Department of Physics and the Center of Theoretical and
Computational Physics, The University of Hong Kong, Pokfulam Road,
Hong Kong, China.\\
$^2$Department of Physics, Beijing Institute of Technology, Beijing
100081, China}

\begin{abstract}
We report a computationally tractable approach to first principles
investigation of time-dependent current of molecular devices under a
step-like pulse. For molecular devices, all the resonant states
below Fermi level contribute to the time-dependent current. Hence
calculation beyond wideband limit must be carried out for a
quantitative analysis of transient dynamics of molecules devices.
Based on the exact non-equilibrium Green's function (NEGF) formalism
of calculating the transient current in Ref.\onlinecite{Maciejko},
we develop two approximate schemes going beyond the wideband limit,
they are all suitable for first principles calculation using the
NEGF combined with density functional theory. Benchmark test has
been done by comparing with the exact solution of a single level
quantum dot system. Good agreement has been reached for two
approximate schemes. As an application, we calculate the transient
current using the first approximated formula with opposite voltage
$V_L(t)=-V_R(t)$ in two molecular structures: Al-${\rm C}_{5}$-Al
and Al-${\rm C}_{60}$-Al. As illustrated in these examples, our
formalism can be easily implemented for real molecular devices.
Importantly, our new formula has captured the essential physics of
dynamical properties of molecular devices and gives the correct
steady state current at $t=0$ and $t\rightarrow \infty$.
\end{abstract}

 \maketitle

\section{introduction}
With the rapid progress in molecular electronics,\cite{Book} quantum
transport in molecular device has received increasing attention. In
particular, the dynamic response of molecular devices to external
parameters\cite{gross3,Zhu,yip,Maciejko,diventra,chen1,gross2}, in
which the external time-dependent fields or internal parametric pump
potentials drive the electrons to tunnel through the molecular
device, is one of the most important issues in molecular
electronics. The simplest molecular device structure is the
two-probe lead-device-lead (LDL) configuration, where ``device" is
the molecular device connected to the external probes by the
``leads". In such a device, all the atomic details of the device
material can be treated using density functional theory (DFT) and
the non-equilibrium physics can be taken into account using
non-equilibrium Green's function (NEGF). Up to now, from an atom
point of view, one of the most popular theoretical approaches used
to study the quantum transport properties of molecular device is
Keldysh nonequilibrium Green's functions coupled with
density-functional theory (NEGF-DFT).\cite{mcdcal} Using this
approach, the steady state quantum transport properties in molecular
devices have been widely studied.

For time dependent response of molecular devices, there have been
many different theoretical approaches, such as evolution of
time-dependent Schrodinger equation,\cite{Schro}, time development
operator approach,\cite{evolution} and the NEGF
technique.\cite{NEGF} These approaches are convenient to deal with
dynamic response of time-dependent external field that is sinusoidal
(e.g., microwave radiation). Under such an external field an
electron can tunnel through the system by emitting or absorbing
photons, giving rise to the photon-assisted tunneling (PAT).
Concerning the steady state ac response to harmonic external field,
the Floquet approach is convenient.\cite{Floquet} For the transient
transport, however, the pulse like ac signal is the optimal driven
force since they can provide a less ambiguous measure of time
scales.\cite{time} In this case, besides PAT, one of the most
interesting questions to ask is how fast a device can turn on or
turn off a current. With the development of molecular electronics,
providing a particular viable switching device has become a key
technical issue. Concerning the transient dynamics, different
approaches such as path-integral techniques,\cite{pathintegral} the
solution of Wigner distribution function,\cite{Wigner} the
time-dependent numerical renormalization group,\cite{NRG}
time-dependent DFT (TDDFT),\cite{TDDFT,gross3} and Keldish Green's
funciton\cite{Zhu,Maciejko,pulse} have also been developed and
applied to different systems. Up to now, most of these approaches
can only be implemented in simple systems such as quantum
dots\cite{Maciejko,pulse} or one-dimension tight-binding
chains.\cite{gross3} Numerical calculation of transient current for
molecular devices is very difficult at present stage due to the huge
computational cost. This is because if we calculate the current as a
function of time $t$, the amount of calculation scales as $t^3$ if
the time-evolution method is used. This scaling can be reduced to a
linear scaling in $t$ if the wideband limit is used.\cite{zheng} As
we have demonstrated,\cite{WangBin} the wideband limit is not a good
approximation for molecular devices. If one uses the exact solution
from NEGF,\cite{Maciejko} one can calculate the transient current at
a particular time. However, the calculation involves a triple
integration over energy which is extremely time consuming. Clearly
an approximate scheme that is suitable for numerical calculation of
transient properties for real molecular devices while still captured
essential physics is needed.

It is the purpose of this paper to provide such a practical scheme.
To study transient dynamics, in this paper, we consider a system
that consists of a scattering region coupled to two leads with the
external time dependent pulse bias potential $V_\alpha(t)=\theta(\pm
t)V_\alpha$. For this case, the time-dependent current for a
step-like pulse has been derived exactly without using the wide-band
limit by Maciejko et al\cite{Maciejko}. Since the general expression
for the current involves triple integrations, it is extremely
difficult to perform them in a real systems like molecules devices.
So, approximation has to be made. The simplest approximation is the
so called wide-band approximation where self-energies $\Sigma^{r,a}$
are assumed to be constants independent of energy.\cite{Jauho1}
Unfortunately, this approximation can not give the correct result
since in general there are several resonant levels that
significantly contribute to the transient current in molecules
devices. To go beyond the wideband limit, we propose an approximate
scheme of calculating the transient current that is suitable for
numerical calculation of real molecules devices.\cite{foot1} Our
scheme is an approximation of the exact solution of Maciejko et
al\cite{Maciejko}. It is very fast computationally and gives the
correct limits at $t=0$ and $t=\infty$. Since the exact solution of
transient current is available for a single level quantum dot
system, we have compared our result with the exact solution on the
quantum dot system to test our approximate schemes. Good agreement
is obtained. Therefore, our approximated scheme maintains essential
physics of transient dynamics. Using our scheme, we calculate the
transient current for the upward pulse (turn-on) and downward pulse
(turn-off) in two molecular structures: Al-${\rm C}_{5}$-Al and
Al-${\rm C}_{60}$-Al. We find that different from the single level
quantum dot system, upon switching on the current oscillates rapidly
in the first a few or tens fs with several characteristic time
scales. Furthermore, due to the resonant states in molecular
devices, transient currents have a much longer decay time $\tau$,
especially for the molecule device having a complex electronic
structure such as Al-${\rm C}_{60}$-Al.

The rest of paper is organized as follows: In Sec.II, starting from
the typical molecular device Hamiltonian which is expressed in an
non-orthogonal basis, we shall derive a general DC and AC current
expressions for a non-orthogonal basis set. It is found that for DC
bias, the expressions of current for orthogonal and non-orthogonal
basis sets are the same. For ac current, however, the expressions
are different as will be demonstrated in Sec.II. The reason that we
study the difference between orthogonal and non-orthogonal basis
sets is the following. For the NEGF formalism, it is assumed that
the basis set is orthogonal. It turns out that for ac transport, the
current expression becomes extremely complicated if non-orthogonal
basis is used. For DFT calculation, however, most people work in
molecular orbitals that are non-orthogonal. Our results show that we
must orthogonalizing the nonorthogonal molecular Hamiltonian, so
that the present approach in Ref.\onlinecite{Maciejko} can be used.
In Sec.III, based on the exact solution of Maciejko et al, we derive
two approximate expressions for transient current with different
levels of approximation. They are all suitable for numerical
calculation for real molecular devices. In addition, the initial
current and its asymptotic long time limit are shown to be correct.
In Sec.IV, in order to appreciate our approximate formulas, we
compare our result with the exact result obtained in
Ref.\onlinecite{Maciejko} for a single-level quantum dot connected
to external leads with a Lorentzian linewidth. In Sec.V, we apply
our formalism to several molecular devices. Finally, a conclusion is
presented in Sec.VI. Two appendices are given at the end of the
paper. In Appendix A, we give a detailed derivation of
orthogonalization relation for an non-orthogonal basis. This
relation is used to derive the effective Green's function which is
the key to approximate exact current expression of Maciejko et al.
In Appendix B, we show how to orthogonalize an nonorthogonal
Hamiltonian so that the general AC current for real molecules device
can be derived.

\section{general AC current}
\subsection{Hamiltonian}
\label{Ham} The transport properties of a molecular device can be
described by the following general Hamiltonian:
\begin{eqnarray}
H=H_c+H_T+\sum_{\alpha=L,R} H_\alpha
\end{eqnarray}
where $H_L$ and $H_R$ describe the left and right macroscopic
reservoir, respectively; $H_c$ is Hamiltonian of the central
molecular device; $H_T$ couples the reservoirs to the molecular
device. For a particular basis set, the above Hamiltonian can be
written in the following matrix form:
\begin{eqnarray}
H_\alpha &=& \sum\limits_{\mu_\alpha\nu_\alpha}
c^{\dagger}_{\mu_\alpha}\left[{\mathbf
H}^0_{\mu_\alpha\nu_\alpha}+eV_{\alpha}(t)\delta_{\mu_\alpha\nu_\alpha}\right]
c_{\nu_\alpha}
\nonumber\\
H_{c}& = &\sum\limits_{\mu_c\nu_c} d^{\dagger}_{\mu_c}\left[
{\mathbf H}^0_{\mu_c\nu_c}+{\mathbf
U}_{\mu_c\nu_c}(t)\right]d_{\nu_c}
\nonumber\\
H_T &=& \sum\limits_{\nu_\alpha,\nu_c}
c^{\dagger}_{\nu_\alpha}{\mathbf T}^0_{\nu_\alpha\nu_c}
d_{\nu_c}+h.c. \label{H}
\end{eqnarray}
where $e$ is the electron charge, $c_{\nu_\alpha}$
($c_{\nu^\dagger_\alpha}$) and $d_{\nu_c}$ ($d^\dagger_{\nu_c}$) are
Fermionic annihilation (creation) operators at the state $\nu$ in
the lead-$\alpha$ and the state $\nu$ in central molecular device.
$\nu_\alpha$, $\nu_c$ are the indices of the given basis set. The
Hamiltonian of lead-$\alpha$ are divided into two parts: the time
independent part ${\bf H}_\alpha^0$ and time dependent part due to
external bias $V_\alpha(t)$ on the lead-$\alpha$. Here we consider
two kinds of step-like bias: upwards pulse (turn-on case)
$V_\alpha^U(t)$ and downwards pulse (turn-off case) $V_\alpha^D(t)$,
where
\begin{eqnarray}
V^D_\alpha(t)=\left\{
\begin{array}{cc}
V_\alpha,~~~ t<0\\
0,~~~~~t>0
\end{array}\right.
,~~~ V^U_\alpha(t)=\left\{
\begin{array}{cc}
0,~~~~~ t<0\\
V_\alpha,~~~t>0
\end{array}\right. \label{bias}
\end{eqnarray}
In the adiabatic approximation it is assumed that the single
particle energies acquire a rigid time-dependent shift as ${\bf
H}_\alpha^0+{\bf I}V_\alpha(t)$. The energy shift in the leads is
assumed to be uniform throughout. This assumption is reasonable
since the pulse rising time is slower than the usual metallic plasma
oscillation time, which ensures that the external electric field is
effectively screened.\cite{Jauho2}

Since Green's function ${\bf G}^r(t,t')$ is obtained by solving
Dyson equation from the known history, it is better to set time
dependent external bias $V_\alpha(t>0)=0$ so that the uncertainty of
future can be eliminated.\cite{ Maciejko} From Eq.(\ref{bias}), this
is satisfied only in the downward case. In the following, we will
discuss how to eliminate this uncertainty for the upward pulse. To
use the Dyson equation, we will separate the Hamiltonian into two
pieces: the unperturbed Hamiltonian that can be exactly resolved and
the interacting term which contributes to the self energy in Dyson
equations. For the downward pulse, we define the non-biased open
system as the unperturbed system. It is described by the Hamiltonian
${\bf H}^{0}={\bf H}^0_\alpha+{\bf H}^0_c+{\bf H}^0_T$. For the
upward pulse, however, the situation is different, in which we will
set the DC biased open system ${\bf H}^V=[{\bf H}^0_\alpha+V_\alpha
{\bf I}]+[{\bf H}^0_c+{\bf U}^V]+{\bf H}^V_T$ as the unperturbed
Hamiltonian and set ${\tilde V}^U_\alpha(t)=V^U_\alpha(t)-V_\alpha$
as the new time dependent part. Here ${\bf H}^V_T$ denotes the
coupling between scattering region and biased leads and ${\bf U}^V$
is the induced coulomb potential due to the external bias. Now, the
time dependent bias ${\tilde V}^U$ satisfies ${\tilde V}^U(t>0)=0$,
and the uncertainty of the future in the upward case is eliminated.
Then, for the downward case, we have ${\tilde
V}^D_\alpha(t)=V^D_\alpha(t)$ and ${\bf H}^{er}={\bf H}^0$ while for
the upward case we have ${\tilde
V}^U_\alpha(t)=V^U_\alpha(t)-V_\alpha$ and ${\bf H}^{ex}={\bf H}^V$.
From now on we will use superscript $``ex"$ to denote the
unperturbed system that is exactly resolvable.

When the system is biased, the incoming electron will polarize the
system. The induced Coulomb potential in the central scattering
region consists of two parts: DC and AC parts. The DC part can be
put into the exactly resolvable Hamiltonian ${\bf H}^{ex}$. The
induced time dependent coulomb potential ${\bf U}(t)$ due to the
external bias ${\tilde V}_\alpha(t)$ is included as part of the
non-equilibrium Hamiltonian. Because the electric field is not
screened in the small scattering region where the potential drop
occurs, the coulomb potential landscape ${\bf U(t)}$ in the central
region is not uniform, which is different from the semi-infinite
leads. Note that it is rather difficult to treat the time-dependent
coulomb potential and no close formed solution exists if one does
not assume wide band limit. In the small bias limit, we can expand
the time-dependent coulomb potential to linear order in bias ${\bf
U}(t)=e\sum_\alpha {\bf u}_\alpha {\tilde V}_\alpha(t)$ so that the
analytic expression for current can be obtained. Here $u_\alpha$ is
the characteristic potential.\cite{buttiker9} From the gauge
invariance, \cite{c2gauge} $\sum_\alpha {\bf u}_\alpha={\bf I}$, and
${\bf u}_\alpha$ is determined from a poisson like
equation.\cite{yadong} In this paper, we consider the symmetric
coupling so that for the external bias ${\tilde V}_L(t)=-{\tilde
V}_R(t)$ it is a good approximation to assume that the time
dependent coulomb potential $U(t)$ is roughly zero in the the
molecular device regime.

In the following, we will derive an exact solution of transient
current using a non-orthogonal basis set.\cite{foot2} To facilitate
the derivation, we take a unitary transformation $\hat{O}(t)$ to the
Hamiltonian (\ref{H}) with
\begin{eqnarray}
\hat{O}(t)&=&{\rm exp}\left\{ie\sum_{\nu_\alpha}\int_0^t d\tau~
\left[{\tilde V}_\alpha(\tau)
c^{\dagger}_{\nu_\alpha}c_{\nu_\alpha}\right]\right\}\nonumber
\end{eqnarray}
where
${\tilde V}_\alpha(\tau)=\theta(-\tau)V_\alpha$ for the downward
pulse and ${\tilde V}_\alpha(\tau)=-\theta(-\tau)V_\alpha$ for the
upward pulse. Note that the time $t$ in $\hat{O}(t)$ can be negative
or positive, and $\hat{O}(t)=1$ only when $t>0$.
The new Hamiltonian ${\mathcal
H}=\hat{O}H\hat{O}^\dagger(t)+i(\frac{\partial}{\partial
t}\hat{O}(t))\hat{O}^\dagger(t)$, in which ${\mathcal H}_\alpha$ and
${\mathcal H}_T$ are different from original ones:
\begin{eqnarray}
{\mathcal H}_\alpha &=& \sum\limits_{\mu_\alpha\nu_\alpha} {\bar
c}^{\dagger}_{\mu_\alpha}{\mathbf H}^0_{\mu_\alpha\nu_\alpha}
{\bar c}_{\nu_\alpha} \nonumber\\
{\mathcal H}_T &=& \sum\limits_{\nu_\alpha,\nu_c} {\bar
c}^{\dagger}_{\nu_\alpha}{\mathbf T}_{\nu_\alpha\nu_c}(t) {
d}_{\nu_c}+h.c. \label{H1}
\end{eqnarray}
where
\begin{eqnarray}
&&{\bar c}_{\nu_\alpha}=c_{\nu_\alpha}
\exp[{ie\sum_{\mu_\alpha}\int_0^t d\tau~{\tilde V}_\alpha(\tau)
{c}^{\dagger}_{\mu_\alpha} {c}_{\mu_\alpha}}],\nonumber \\
&&{\mathbf T}_{\nu_\alpha \nu_c}(t)={\mathbf T}^0_{\nu_\alpha
\nu_c}{\mathfrak W}_\alpha(t)\nonumber \\
&&{\mathfrak W}_\alpha(t)=\exp[{ie\int_0^t {\tilde
V}_\alpha(\tau)d\tau}]\label{H2}
\end{eqnarray}
For the original Hamiltonian with nonorthogonal basis, the overlap
between nonorthogonal basis is expressed as the matrix form ${\bf
S}^0_{\mu\nu}=\langle\mu|\nu\rangle$. After the unitary transform,
annihilation (creation) operators $c_\alpha$ ($c^\dagger_\alpha$)
and consequently the orbital basis $\mu_\alpha$ in the leads are
changed, then overlap matrices between the leads and the scattering
region become
\begin{eqnarray}
&&{\bf S}_{\nu_\alpha \nu_c}(t)={\bf S}^0_{\nu_\alpha \nu_c}
{\mathfrak W}_\alpha(t)\nonumber \\
&&{\bf S}_{\nu_c\nu_\alpha}(t)={\mathfrak W}^\dagger_\alpha(t){\bf
S}^0_{\nu_c\nu_\alpha}.\label{H3}
\end{eqnarray}
In the following, we will use the transformed Hamiltonian
[Eq.(\ref{H1},\ref{H2}), in which ${\bar c}_{\nu_\alpha}$, ${
d}_{\nu_c}$ are used] to derive the time dependent current
expression.

\subsection{The current}
The current operator from a particular lead-$\alpha$ to the
molecular junction can be calculated from the evolution of the
number operator of the electron in the semi-infinite lead-$\alpha$.
Assuming there is no direct coupling between the left and right
leads, the current operator can be expressed as:\cite{acCur}
\begin{eqnarray}
\hat{J}_\alpha(t)&=&-e\sum_{\nu_\alpha}\frac{d}{dt}\hat{N}_{\nu_\alpha}(t)\nonumber \\
&=&-e\sum_{\nu_\alpha}\left[{\bar c}^\dagger_{\nu_\alpha}(t)
\frac{d}{dt}{\bar c}_{\nu_\alpha}(t)+\left(\frac{d}{dt}{\bar
c}^\dagger_{\nu_\alpha}(t)\right)
{\bar c}_{\nu_\alpha}(t)\right]\nonumber \\
&=&e\sum_{\nu_\alpha,\nu_c}{\bar
c}^\dagger_{\nu_\alpha}(t)\left(i{\mathbf
T}_{\nu_\alpha\nu_c}(t)+{\mathbf
S}_{\nu_\alpha\nu_c}(t)\frac{d}{dt}\right)d_{\nu_c}(t)+H.c.\label{Cur1}
\end{eqnarray}
where `H.c.' denotes the Hermitian conjugate. The current is
obtained by taking average over the nonequilibrium quantum state
`$<...>$',
\begin{eqnarray}
J_\alpha(t)=e~\sum_{\nu_\alpha,\nu_c} &&\left[\mathbf{G}^<_{\nu_c
\nu_\alpha}(t,t')\left(\mathbf{T}_{\nu_\alpha,\nu_c}(t')-{\bf
S}_{\nu_\alpha,\nu_c}(t')i\grave{\frac{\partial}{\partial
t}}\right)\right.\nonumber\\
-&&\left.\left(\mathbf{T}_{\nu_c,\nu_\alpha}(t')-{\bf
S}_{\nu_c,\nu_\alpha}(t')i\acute{\frac{\partial}{\partial
t}}\right)\mathbf{G}^<_{\nu_\alpha\nu_c}(t',t)\right]_{t=t'},
\label{Cur2}
\end{eqnarray}
where ``$\grave{\frac{\partial}{\partial t}}$" and
``$\acute{\frac{\partial}{\partial t}}$" denotes the left and right
derivation respectively, and $${\bf G}^<_{\nu_c,\nu_\alpha}(t,t')=i
\left\langle {\bar
c}^\dagger_{\nu_\alpha}(t')d_{\nu_c}(t)\right\rangle,~~ {\bf
G}^<_{\nu_\alpha,\nu_c}(t',t)=i\left\langle
d^\dagger_{\nu_c}(t){\bar c}_{\nu_\alpha}(t')\right\rangle.$$

Using the Keldysh equation and the theorem of analytic continuation,
we have
\begin{eqnarray}
{\bf G}^<_{c\alpha}(t,t')=\int dt_1&&\left[{\bf
G}^r_{cc}(t,t_1){\bf B}_{c\alpha}(t_1){\bf g}^<_{\alpha\alpha}(t_1,t')+\right.\nonumber \\
&&\left.{\bf G}^<_{cc}(t,t_1){\bf B}_{c\alpha}(t_1){\bf
g}^a_{\alpha\alpha}(t_1,t')\right]\label{Cur3}
\end{eqnarray}
where
\begin{eqnarray}
{\bf B}_{c\alpha}(t_1)&=&{\bf T}_{c\alpha}(t_1)-{\bf S
}_{c\alpha}(t_1)i\grave{\frac{\partial}{\partial t}}\label{B}
\end{eqnarray}
For simplicity, we have dropped the subscript $\mu$, and keep only
the symbol $c$ and $\alpha$ to indicate the central scattering
region and lead-$\alpha$, respectively. In the above expression and
in the following, the summation convention on repeated sub-indices
is assumed. Substituting Eq.(\ref{Cur3}) into Eq.(\ref{Cur2}), we
have the general expression for the current:
\begin{eqnarray}
J_\alpha(t)&=&-2e{\rm Re}\int dt_1~{\rm Tr}
\nonumber\\
&&\left[{\bf G}^r_{cc}(t,t_1){\bf B}_{c\alpha}(t_1){\bf
g}^<_{\alpha\alpha}(t_1,t'){\bf B}_{\alpha c}(t')-\right.\nonumber\\
&&\left.{\bf G}^<_{cc}(t,t_1){\bf B}_{c\alpha}(t_1){\bf
g}^a_{\alpha\alpha}(t_1,t'){\bf B}_{\alpha
c}(t')\right]_{t=t'}\label{Cur4}
\end{eqnarray}

When the system reaches a stationary state, $V_\alpha(t)=V_\alpha$
becomes time independent, from definition Eq.(\ref{H2}), (\ref{H3})
and (\ref{B}), we can find
$${\bf B}_{c\alpha}(t_1)X{\bf B}_{\alpha c}(t)=
e^{-ieV_\alpha(t_1-t)}{\bf B}^0_{c\alpha}X{\bf B}^0_{\alpha c},$$
with ${\bf B}^0_{c\alpha/\alpha c} = {\bf T}^0_{c\alpha/\alpha
c}-i\grave{\frac{\partial}{\partial t}}{\bf S }^0_{c\alpha/\alpha
c}$, where ``0" denotes the zero bias system.In addition, all the
propagators ${\bf G}$ and ${\bf g}$ depend only on the time
difference $t_1-t$. Taking the Fourier transformation, from
Eq.(\ref{Cur2}) or Eq.(\ref{Cur4}), we can easily obtain DC current
expressed in the energy representation:
\begin{eqnarray}
J_\alpha&=&\int d\epsilon ~\mathcal{J}_{\alpha}(\epsilon)\nonumber
\\&=&\rm{Re}~2e\int d\epsilon~{\rm
Tr}\left[\mathbf{G}^r(\epsilon)\mathbf{\Sigma}^{<}_\alpha(\epsilon)+
\mathbf{G}^<(\epsilon)\mathbf{\Sigma}^{a}_\alpha(\epsilon)\right]\label{dcCur}
\end{eqnarray}
where ${\mathbf G}$ and ${\mathbf \Sigma}$ are the Green's function
and the self-energy. They have the same matrix dimension as that of
the Hamiltonian ${\bf H}_c$. The Green's function ${\bf G}^{r/a}$
and self-energy ${\bf \Sigma}^{r/a}$ is defined as
\begin{eqnarray}
&&{\bf G}^{r/a}(\epsilon)=\left[\epsilon{\bf I}-{\bf H}_c-{\bf
\Sigma}^{r/a}(\epsilon)\right]^{-1}\nonumber \\
&&\mathbf{\Sigma}^{\gamma}_\alpha(\epsilon)=
\left[\mathbf{T}^{0}_{c\alpha}-\epsilon^\alpha{\bf
S}^0_{c\alpha}\right]\mathbf{g}^{\gamma}_{\alpha\alpha}(\epsilon^\alpha)
\left[\mathbf{T}^0_{\alpha c}-\epsilon^\alpha{\bf S}^0_{\alpha
c}\right]\label{dcself}
\end{eqnarray}
where $\epsilon^\alpha=\epsilon-eV_\alpha$, ${\bf I}$ is the unitary
matrix with same dimension as ${\bf H}_c$, $\gamma=r,a,<$, and
\begin{eqnarray}
{\bf g}^{r/a}_{\alpha\alpha}(\epsilon)&=&\left[\left((\epsilon\pm
i0^+){\bf S}^0_{\alpha\alpha}-{\bf
H}^0_{\alpha\alpha}\right)^{-1}\right]_{\nu_\alpha \in {\rm sur},
\mu_\alpha \in {\rm sur}}\nonumber \\
{\bf g}^{<}_{\alpha\alpha}(\epsilon)&=&f(\epsilon)\left[{\bf
g}^{a}_{\alpha\alpha}(\epsilon)-{\bf
g}^{r}_{\alpha\alpha}(\epsilon)\right]\label{surfself}
\end{eqnarray}
is the surface Green's function of the semi-infinite periodic lead
which can be calculated numerically using a transfer matrix
method.\cite{transfer} Here, $f(\epsilon)$ is the Fermi
distribution.
Eq.(\ref{dcCur}) shows that the dc current expressions are the same
for both orthogonal and non-orthogonal basis sets.

When the time dependent field $V_\alpha(t)$ is present, however, the
current expressed in energy representation will be very complicated
for nonorthogonal basis due to the term ${\bf S
}(t')i\frac{\partial}{\partial t}$ in Eq.(\ref{Cur2}), since ${\bf
B}(t_1)X{\bf B}(t)$ can't be expressed as a function of time
difference $t_1-t$. One thing is clear, the transient current
expressions are different for orthogonal and non-orthogonal basis
sets. Instead of deriving a complicated transient current expression
using a non-orthogonal basis set, we will eliminate ${\bf
S}_{c\alpha/\alpha c}(t')i\frac{\partial}{\partial t}$ in
Eq.(\ref{Cur2}) and work on an orthogonal basis set. In Appendix
\ref{Orbasis}, from the overlap matrix ${\bf S}$, we derive the
orthogonal basis set and new Hamiltonian ${\tilde H}$ expressed in
this orthogonal basis. With the new orthogonal Hamiltonian, the
overlap matrix ${\bf S}_{c\alpha/\alpha c}(t')$ will be eliminated
since the overlap matrix of orthogonal basis ${\bf S}^{orth}={\bf
I}$. Then, replacing Hamiltonian ${\bf H}$ in Eq.(\ref{H}) with
$\tilde{\bf H}$ and go through the derivation leading to
Eqs.(\ref{H}-\ref{Cur4}) again, we arrive at a new AC current
expression:
\begin{eqnarray}
&&J_\alpha(t)=2e{\rm Re}\int dt_1{\rm Tr}\left\{
\mathbf{G}_{cc}^r(t,t_1)\left[{\bf T}_{c\alpha}(t_1){\bf
g}^{<,ex}_{\alpha\alpha}(t_1-t){\bf T}_{\alpha
c}(t)\right]\right\}\nonumber
\\&&+2e{\rm Re}\int dt_1{\rm Tr}\left\{
\mathbf{G}^<_{cc}(t,t_1)\left[{\bf T}_{c\alpha}(t_1){\bf
g}^{a,ex}_{\alpha\alpha}(t_1-t){\bf T}_{\alpha
c}(t)\right]\right\}\label{Cur5}
\end{eqnarray}
Defining the self-energy on the orthogonal basis
\begin{eqnarray}
{\bf \Sigma}^{\gamma=r,a,<}_\alpha(t,t')={\bf T}_{c\alpha}(t){\bf
g}^{\gamma,ex}_{\alpha\alpha}(t-t'){\bf T}_{\alpha
c}(t')\label{self1}
\end{eqnarray}
where ${\bf g}^{\gamma,ex}_{\alpha\alpha}(t-t')=\int
\frac{d\epsilon}{2\pi} ~e^{-i\epsilon (t-t')}{\bf
g}^{\gamma,ex}_{\alpha\alpha}(\epsilon)$ is the surface Green's
function of semi-infinite lead-$\alpha$ in the unperturbed state as
defined in the Sec.\ref{Ham}. For the downward pulse we have set the
unperturbed system as the open system at zero bias, in which ${\bf
g}^{\gamma,ex}_{\alpha\alpha}(\epsilon)=\left[\epsilon-H^0_\alpha+i0^+\right]^{-1}_{\alpha\in{\rm
sur}}$. For the upward pulse, the unperturbed system means
$V_\alpha$ biased open system, in which ${\bf
g}^{\gamma,eq}_{\alpha\alpha}(\epsilon)=\left[\epsilon-eV_\alpha-H^0_\alpha+i0^+\right]^{-1}_{\alpha\in{\rm
sur}}$. From Eq.(\ref{Cur5}),(\ref{self1}), we have the general
current formula
\begin{eqnarray}
J_\alpha(t)&=&2e{\rm Re}\int dt_1 {\rm
Tr}\left[\mathbf{G}^r(t,t_1)\Sigma^<_\alpha(t_1,t) +
\mathbf{G}^<(t,t_1)\Sigma^a_\alpha(t_1,t)\right]\nonumber \\
\label{generalCur}
\end{eqnarray}

At $t<0$, AC external bias $V_\alpha(t)$ or time dependent part in
Hamiltonian ${\tilde V}_\alpha(t)$ is a constant and the system is
in a steady state. Consequently, the total current is known from DC
transport theory that is expressed in the form of Eq.(\ref{dcCur})
but with the Green's function and self-energy obtained from the
orthogonal Hamiltonian defined above. Hence in the following we
shall derive only the Ac current when $t>0$. First, we shall look at
the self-energy. From Eq.(\ref{H2}) and (\ref{self1}),
\begin{eqnarray}
\mathbf{\Sigma}^{\gamma}_\alpha(t,t')&=&\mathfrak{W}^\dagger_\alpha(t)
\left[\mathbf{T}^{0}_{c\alpha}\mathbf{g}_{\alpha\alpha}^{\gamma}(t,t')T^0_{\alpha
c}\right]\mathfrak{W}_\alpha(t')\nonumber \\
&=&\mathfrak{W}^\dagger_\alpha(t)\left[\int \frac{d\epsilon}{2\pi}
~e^{i\epsilon(t-t')}\mathbf{\Sigma}_{\alpha}^{\gamma,ex}(\epsilon)\right]
\mathfrak{W}_\alpha(t')\nonumber \\
&=&\mathfrak{W}^\dagger_\alpha(t)\mathfrak{V}^\dagger_\alpha(t)
\left[\int \frac{d\epsilon}{2\pi}
~e^{i\epsilon(t-t')}\mathbf{\Sigma}_{\alpha}^{\gamma,0}(\epsilon)\right]
\mathfrak{V}_\alpha(t')\mathfrak{W}_\alpha(t') \nonumber
\\
\label{self2}
\end{eqnarray}
where $\mathfrak{V}_\alpha(t)=1$ for the downward pulse and $
\mathfrak{V}_\alpha(t)=e^{ie V_\alpha t}$ for the upward pulse. Here
$\mathbf{\Sigma}_{\alpha}^{\gamma,0}(\epsilon)$ is the self-energy
at zero bias,
$\mathbf{\Sigma}_{\alpha}^{\gamma,ex}(\epsilon)=\mathbf{T}^{0}_{c\alpha}
\mathbf{g}_{\alpha\alpha}^{\gamma,ex}(\epsilon)\mathbf{T}^{0}_{\alpha
c}$ is the self-energy at the unperturbed state defined above. In
the downward case ${\bf \Sigma}^{\gamma,ex}_\alpha={\bf
\Sigma}^{\gamma,0}_\alpha$; In the upward case ${\bf
\Sigma}^{\gamma,ex}_\alpha={\bf \Sigma}^{\gamma,V}_\alpha$. Setting
${\bf S}^0_{\alpha c}={\bf S}^0_{c\alpha}=0$, ${\bf
\Sigma}^{\gamma,0}_\alpha$ and ${\bf \Sigma}^{\gamma,V}_\alpha$ are
defined in Eq.(\ref{dcself}) with zero and nonzero $V_\alpha$,
respectively. We have ${\bf \Sigma}^{r/a,V}_\alpha(\epsilon)={\bf
\Sigma}^{r/a,0}_\alpha(\epsilon-eV_\alpha)$. From
Eq.(\ref{generalCur}) and (\ref{self2}), we find
\begin{eqnarray}
J_{\alpha}(t)&=&2e\rm{Re}\int\frac{d\epsilon}{2\pi} \int^t_{-\infty}
dt_1~~e^{i\epsilon(t-t_1)}\nonumber \\
&&\left[\mathbf{G}^r(t,t_1)
\tilde{\mathbf{\Sigma}}^<_\alpha(\epsilon,t_1,t)
+\mathbf{G}^<(t,t_1)\tilde{\mathbf{\Sigma}}^a_\alpha(\epsilon,t_1,t)\right]\label{Cur6}
\end{eqnarray}
where the first term is the current flowing into the molecular
device while the second one is the current flowing from the
molecular device, and
\begin{eqnarray}
\tilde{\mathbf{\Sigma}}^{\gamma}_\alpha(\epsilon,t_1,t)
&=&\mathcal{W}_\alpha^\dagger(t_1)
\mathbf{\Sigma}^{\gamma,0}_\alpha(\epsilon)\mathcal{W}_\alpha(t)
\label{self3}
\end{eqnarray}
where
$\mathcal{W}_\alpha(t)=\mathfrak{V}_\alpha(t)\mathfrak{W}_\alpha(t)$.
Here ${\bf \Sigma}^{\gamma,0}_{\alpha\alpha}$ is the self-energy of
lead-$\alpha$ at zero bias. The lesser Green's function is given by
\begin{eqnarray}
&&\mathbf{G}^<(t,t')=\int dt_1 \int dt_2~\mathbf{G}^r(t,t_1)
\left[\sum_\beta\mathbf{\bf{\Sigma}}^{<}_\beta(t_1,t_2)\right]\mathbf{G}^a(t_2,t')
\nonumber \\
&=& \sum_\beta \int \frac{d\epsilon}{2\pi}~ e^{-i\epsilon(t-t')}
\left[\int^t_{-\infty}
dt_1~e^{i\epsilon(t-t_1)}\mathcal{W}_\beta(t)\mathbf{G}^r(t,t_1)\mathcal{W}^\dagger_\beta(t_1)\right]
\nonumber \\
&&\mathbf{\Sigma}^{<,0}_\beta(\epsilon)\left[\int^{t'}_{-\infty}
dt_2~e^{-i\epsilon(t'-t_2)}\mathcal{W}_\beta(t_2)\mathbf{G}^a(t',t_2)
\mathcal{W}^\dagger_\beta(t)\right]
\label{Gl}
\end{eqnarray}
Substitute Eq.(\ref{self3}) and (\ref{Gl}) into Eq.(\ref{Cur6}) and
introducing a spectrum function
\begin{eqnarray}
{\bf A}_\alpha(t,\epsilon)=\int_{-\infty}^t dt_1~
e^{i\epsilon(t-t_1)}\mathcal{W}_\alpha(t){\bf
G}^r(t,t_1)\mathcal{W}^\dagger_\alpha(t_1)\label{A1}
\end{eqnarray}
we have
\begin{eqnarray}
J^{in}_{\alpha}(t)&=&2e\rm{Re}\int \frac{d\epsilon}{2\pi}~{\bf
A}_\alpha(t,\epsilon){\bf
\Sigma}_\alpha^{<,0}(\epsilon)\label{Jin} \\
J^{out}_{\alpha}(t)&=&2e\rm{Re}\int
\frac{d\epsilon}{2\pi}~\sum_\beta{\bf A}_\beta(t,\epsilon){\bf
\Sigma}^{<,0}_\beta(\epsilon)\tilde{\bf
F}_{\beta\alpha}(t,\epsilon)\label{Cur7}
\end{eqnarray}
where
\begin{eqnarray}
\tilde{\bf F}_{\beta\alpha}(t,\epsilon)&=&\int_{-\infty}^t
dt'~e^{-i\epsilon(t-t')}\int \frac{dE}{2\pi}~e^{iE(t-t')}\nonumber\\
&& {\bf A}_\beta^\dagger(t',\epsilon){\mathcal W}_\alpha^\dagger(t')
{\bf \Sigma}_\alpha^{a,0}(E){\mathcal W}_\alpha(t)\label{F1}
\end{eqnarray}

Very often, ${\bf \Sigma}^{r/a}(t-t')$ is singular at $t=t'$, such
as the quantum dot system with the wide-band limit ${\bf
\Sigma}^{r/a}(0)=\int \frac{dE}{2\pi}~ {\bf
\Sigma}^{r/a}(E)=\delta(0)(\mp \Gamma/2)$, or the
superconducting-quantum dot-normal metal system, and so on. In these
cases, we should be careful with Eq.(\ref{F1}),
\begin{eqnarray}
\tilde{\bf F}_{\beta\alpha}(t,\epsilon)&=&{\bf
F}_{\beta\alpha}(t,\epsilon)+\bar{\bf
F}_{\beta\alpha}(t,\epsilon)\nonumber\\
&=&\left(\int_{-\infty}^{t^-}+\frac{1}{2}\int_{t^-}^{t^+}\right)
dt'~e^{-i\epsilon(t-t')}\int \frac{dE}{2\pi}~e^{iE(t-t')}\nonumber\\
&& {\bf A}_\beta^\dagger(t',\epsilon){\mathcal W}_\alpha^\dagger(t')
{\bf \Sigma}_\alpha^{a,0}(E){\mathcal W}_\alpha(t)\label{F2}
\end{eqnarray}
The first integral $\int_{-\infty}^{t^-}$ is the same as
Eq.(\ref{F1}), the second integral $\frac{1}{2}\int_{t^-}^{t^+}$ now
becomes $ \bar{\bf F}_{\beta\alpha}(t,\epsilon)= {\bf
A}_\beta^\dagger(t,\epsilon) {\bf \Delta}^{a}_\alpha$, where we have
defined
\begin{eqnarray}
{\bf \Delta}^{r/a}_\alpha&=&\frac{1}{2}\int_{t^-}^{t^+}
dt'~\left[\int
\frac{dE}{2\pi}~ {\bf \Sigma}_\alpha^{r/a,0}(E)\right]\nonumber \\
&=&\frac{1}{2}\int_{t^-}^{t^+} dt'~{\bf \Sigma}_\alpha^{r/a,0}(0)
\end{eqnarray}
Then, Eq.(\ref{Cur7}) becomes
\begin{eqnarray}
J^{out}_{\alpha}(t)&=&2e\rm{Re}\int
\frac{d\epsilon}{2\pi}~\sum_\beta{\bf A}_\beta(t,\epsilon){\bf
\Sigma}^{<,0}_\beta(\epsilon){\bf
F}_{\beta\alpha}(t,\epsilon)\nonumber \\
&+&2e\rm{Re}\int \frac{d\epsilon}{2\pi}~\sum_\beta{\bf
A}_\beta(t,\epsilon){\bf \Sigma}^{<,0}_\beta(\epsilon){\bf
A}^\dagger_{\beta}(t,\epsilon){\bf \Delta}^{a}_\alpha\label{Jout}
\end{eqnarray}
We note that Eq.(\ref{Jout}) is the same as that derived in
Ref.\onlinecite{Maciejko}. Different from Ref.\onlinecite{Maciejko},
we have split the expression into two terms. The first term
corresponds to the non-wideband limit, i.e., when the linewidth
function $\bf \Gamma$ goes to zero at large energy. The second term
of Eq.(\ref{Jout}) is related to the wideband limit. Hence, for a
quantum dot with a Lorentzian linewidth function\cite{Maciejko},
only the first term is nonzero while for the system in contact with
a superconducting lead both terms are nonzero.

So far, we have discussed the ac {\it conduction} current
$J_\alpha(t)$ under the time dependent bias derived from the
evolution of the number operator of the electron in the
semi-infinite lead-$\alpha$. Now we wish to address the issue of
charge accumulation in the scattering region. In principle, this can
be done by including the self-consistent Coulomb potential due to ac
bias.\cite{yadong} However, at finite voltages, there is no close
form expression for ac current if Coulomb potential is included.
Alternatively, one can treat Coulomb potential phenomenologically as
follows. From the continuity equation, $\sum_\alpha J_\alpha(t)
+dQ(t)/dt=0$, we see that the conduction current is not a conserved
quantity. In the presence of ac bias, the displacement current
$J_\alpha^d$ due to the charge pileup $dQ/dt$ inside the scattering
region becomes important and must be considered. Since we have
neglected the Coulomb interaction in our calculation, we can use the
method of current partition\cite{buttiker4,wbg} to include the
displacement current. This can be done by partitioning the total
displacement current $\sum_\alpha J^d_\alpha=dQ/dt$ into each leads
giving rise to a conserving total current
$I_\alpha=J_\alpha+J^d_\alpha$. For symmetric systems like what we
shall study below, it is reasonable to assume that $J^d_L=J^d_R$
from which we find $J^d_\alpha=-(J_L+J_R)/2$. Hence the total
current is given by $I_L = (J_L-J_R)/2$\cite{Jauho2} which satisfies
the current conservation $I_L+I_R=0$.

\section{transient AC current}

Up to now, we have derived the general expression for time dependent
current, Eq.(\ref{A1},\ref{Jin},\ref{F1},\ref{Jout}) which can be
used for orthogonal as well as nonorthogonal basis set. To calculate
the transient current we have to solve the retarded Green's function
${\bf G}^r(t,t')$ and integrate it over time to find ${\bf
A}_{\beta}(t,\epsilon)$ and $\tilde{\bf
F}_{\beta\alpha}(t,\epsilon)$. For the pulse-like voltage ${\tilde
V}_\alpha(t)=\pm \theta(-t)$, we can obtain the Green' function
${\bf G}^r(t,t')$ by solving Dyson equation ${\bf G}^r={\bf
G}^{r,eq}+{\bf G}^{r,eq}{\bf \Xi}{\bf G}^{r}$ from the known history
in the time domain. Depending on what is the chosen unperturbed
system that can be solved exactly, the Dyson equation can be written
in a different but equivalent form. In the study of time-dependent
transport, it is better to treat the time-independent, open steady
state system as the unperturbed system as described in
Sec.\ref{Ham}, and treat the time dependent part
$\tilde{V}_\alpha(t)$ and ${\bf U}(t)$ as a perturbation. As a
result, the effective self-energy ${\bf \Xi}$, which is due to the
ac bias, would have two sources: the perturbation in leads $\bar{\bf
\Sigma}^r_\alpha$ and the induced Coulomb interaction in molecular
device ${\bf U}(t)$. Then,
\begin{eqnarray}
\mathbf{G}^r(t,t')&=&\mathbf{G}^{r,ex}(t,t')+\int_{-\infty}^0
dt_1~\mathbf{G}^{r,ex}(t,t_1) \mathbf{U}(t_1)\mathbf{G}^r(t_1,t')
\nonumber
\\&+&~\int dt_1 ~ dt_2~\mathbf{G}^{r,ex}(t,t_1)
\left[\sum_\alpha\mathbf{\bar{\Sigma}}^{r}_\alpha(t_1,t_2)\right]
\mathbf{G}^r(t_2,t') \nonumber
\end{eqnarray}
where ${\bf U}(t)$ is the response of the molecular device that is
due to the Coulomb interaction when the time-dependent voltage is
turned on. Here we have assumed an adiabatic response since most of
time the variance of the applied electric field is much slower than
the particles' intrinsic lifetime inside the scattering region. Then
we have ${\bf U}(t)=\pm{\bf U}\theta(-t)$ for downward case and
upward case with ${\bf U}={\bf H}^V_c-{\bf H}^0_c$.
$$\int dt_1~dt_2=\left(\int^0_{-\infty} dt_1 \int^{t_1}_{-\infty}
dt_2 + \int^t_0 dt_1 \int^0_{-\infty} dt_2\right)$$
\begin{eqnarray}
&&\mathfrak{\bar{{\bf \Sigma}}}^{r}_{\alpha}(t,t')={\bf
\Sigma}^r_\alpha(t,t')- {\bf \Sigma}_\alpha^{r,ex}(t-t')\nonumber\\
&&{\bf \Sigma}_\alpha^{r,ex}(t-t')={\mathfrak V}^\dagger_\alpha(t)
{\bf \Sigma}_\alpha^{r,0}(t-t'){\mathfrak V}_\alpha(t')\nonumber
\end{eqnarray}

\subsection{Exact expression of ${\bf A}_\beta(t,\epsilon)$ and ${\bf
F}_{\beta\alpha}(t,\epsilon)$}

Following the derivations in Ref.\onlinecite{Maciejko}, we can get
the exact expression for ${\bf A}_\beta(t,\epsilon)$ and ${\bf
F}_{\beta\alpha}(t,\epsilon)$ with the aid of the expressions
$\epsilon_\beta=\epsilon+eV_\beta$ and
$\epsilon_{\beta\alpha}=\epsilon+eV_\beta-eV_\alpha$:
\begin{eqnarray}
{\bf A}^D_\beta(t,\epsilon)&=&{\bf G}^{r,0}(\epsilon)
+\int\frac{dE}{2\pi}~e^{i(\epsilon-E)t}\nonumber \\
&\times& {\bf G}^{r,0}(E)\left[Z(\epsilon_\beta)- Z(\epsilon)+{\bf
P}_D{\bf
G}^{r,V}(\epsilon_\beta)\right] \label{AD1}\\
{\bf
F}^D_{\beta\alpha}(t,\epsilon)&=&\int\frac{dE}{2\pi}~Z^*(\epsilon){\bf
G}^{a,0}(\epsilon){\bf
\Sigma}^{a,0}_\alpha(E)+\int\frac{dE}{2\pi}~e^{-i(\epsilon-E)t}
\nonumber\\
&\times&\left\{\left[Z^*(\epsilon_\beta)-Z^*(\epsilon)+ {\bf
G}^{a,V}(\epsilon_\beta){\bf P}^\dagger_D\right] {\bf
G}^{a,0}(E){\bf Q}_D(E)\right.\nonumber
\\&+& \left.\left[Z^*(\epsilon_{\beta\alpha}){\bf
G}^{a,V}(\epsilon_\beta)-Z^*(\epsilon){\bf
G}^{a,0}(\epsilon)\right]{\bf \Sigma}^{a,0}_\alpha(E)
\right\}\label{FD1}
\end{eqnarray}
\begin{eqnarray}
{\bf A}^U_\beta(t,\epsilon)&=&{\bf
G}^{r,V}(\epsilon_\beta)+\int\frac{dE}{2\pi}~
e^{i(\epsilon_\beta-E)t}\nonumber\\&\times&{\bf
G}^{r,V}(E)\left[Z(\epsilon)-Z(\epsilon_\beta)+{\bf P}_U{\bf
G}^{r,0}(\epsilon)\right]
\label{AU1}\\
{\bf F}^U_{\beta\alpha}(t,\epsilon)&=&\int\frac{dE}{2\pi}~
Z^*(\epsilon_{\beta\alpha}){\bf G}^{a,V}(\epsilon_\beta){\bf
\Sigma}^{a,0}_\alpha(E)+\int\frac{dE}{2\pi}~e^{-i(\epsilon_\beta-E)t}
\nonumber\\
&\times&\left\{\left[Z^*(\epsilon)-Z^*(\epsilon_\beta)+ {\bf
G}^{a,0}(\epsilon){\bf P}^\dagger_U\right] {\bf G}^{a,V}(E){\bf
Q}_U(E)\right.\nonumber
\\&+& \left.e^{ieV_\alpha t}\left[Z^*(\epsilon){\bf
G}^{a,0}(\epsilon)-Z^*(\epsilon_{\beta\alpha}){\bf
G}^{a,V}(\epsilon_\beta)\right]{\bf \Sigma}^{a,0}_\alpha(E)
\right\}\nonumber \\
&&\label{FU1}
\end{eqnarray}
where
\begin{eqnarray}
{\bf P}_D&=&Z(\epsilon_\beta){\bf U} +\sum_\delta
\left[Z(\epsilon_\beta)-Z(\epsilon_{\beta\delta})\right] \left[{\bf
\Sigma}^{r,0}_\delta(\epsilon_{\beta\delta})
-{\bf \Sigma}^{r,0}_\delta(E)\right]\nonumber\\
{\bf P}_U&=&-Z(\epsilon){\bf U}+\sum_\delta
\left[Z(\epsilon)-Z(\epsilon_\delta)\right] \left[{\bf
\Sigma}^{r,0}_\delta(\epsilon)- {\bf
\Sigma}^{r,0}_\delta(E-V_\delta)\right]\nonumber \\
&&{\bf Q}_D(E)=\int
\frac{d\epsilon'}{2\pi}~\left[1-e^{i(\epsilon'-E)t}\right]
Z(\epsilon'){\bf \Sigma}^{a,0}_\alpha(\epsilon')\nonumber \\
&&{\bf Q}_U(E)=\int
\frac{d\epsilon'}{2\pi}~\left[1-e^{i(\epsilon'_\alpha-E)t}\right]
Z(\epsilon'_\alpha){\bf \Sigma}^{a,0}_\alpha(\epsilon') 
\label{expc}
\end{eqnarray}
with
\begin{eqnarray}
Z(\epsilon)=[i(E-\epsilon-i0^+)]^{-1} \label{zz}
\end{eqnarray}
In the absence of the ac bias, the quantity $A_\alpha$ is the
Fourier transform of the retarded Green's function while the
quantity $F_{\beta \alpha}$ is related to the Fourier transform of
the advanced Green's function. They are all expressed in terms of
the unperturbed Green's functions ${\bf G}^{r/a,0/V}$ and self
energy ${\bf \Sigma}^{0/V}$ which have been widely studied in
molecular device using the NEGF-DFT formalism. ${\bf G}^{r/a,0/V}$
and self energy ${\bf \Sigma}^{0/V}$ can be expressed as
\begin{eqnarray}
&&{\bf G}^{r/a,0/V}(\epsilon)=\left[\epsilon{\bf I}-{\bf
H}^{0/V}_c-{\bf
\Sigma}^{r/a,0/V}(\epsilon)\right]^{-1}\nonumber \\
&&\mathbf{\Sigma}^{\gamma,0}_\alpha(\epsilon)=
\left[\mathbf{T}^{0}_{c\alpha}-\epsilon{\bf
S}^0_{c\alpha}\right]\mathbf{g}^{\gamma}_{\alpha\alpha}(\epsilon)
\left[\mathbf{T}^0_{\alpha c}-\epsilon{\bf
S}^0_{\alpha c}\right]\nonumber\\
&&\mathbf{\Sigma}^{\gamma,V}_\alpha(\epsilon)=
\left[\mathbf{T}^{0}_{c\alpha}-\epsilon^\alpha{\bf
S}^0_{c\alpha}\right]\mathbf{g}^{\gamma}_{\alpha\alpha}(\epsilon^\alpha)
\left[\mathbf{T}^0_{\alpha c}-\epsilon^\alpha{\bf S}^0_{\alpha
c}\right]\nonumber
\end{eqnarray}
where $\gamma=r,a,<$, $\epsilon^\alpha=\epsilon-eV_\alpha$.
Obviously, ${\bf \Sigma}^{\gamma,V}_\alpha(\epsilon)={\bf
\Sigma}^{\gamma,0}_\alpha(\epsilon-eV_\alpha)$. In the wideband
limit, Eq.(\ref{AD1}-\ref{FU1}) will reduce to the formula first
derived by Jauho et al.\cite{Jauho2} With ${\bf A}$ and ${\bf F}$
obtained we can, in principle, solve the AC current biased by
downwards or upwards pulse exactly. In practice, however, its
computational cost is expensive for a realistic molecular device.
For example, to calculate $J_\alpha^{out}(t)$, we have to do triple
integrals over energy and repeat this procedure to collect data for
all time sequence. In the numerical calculation especially in
ab-initio modeling, it is practically very difficult if not
impossible to calculate the transient current for the complex
structure in molecular devices. So approximation must be made so
that Eq.(\ref{AD1}-\ref{FU1}) can be simplified.

\subsection{Approximate scheme of ${\bf A}_\beta(t,\epsilon)$ and ${\bf
F}_{\beta\alpha}(t,\epsilon)$}

The approximate solution of ${\bf A}_\beta(t,\epsilon)$ and ${\bf
F}_{\beta\alpha}(t,\epsilon)$ in Eq.(\ref{AD1}-\ref{FU1}) have to
satisfy the following requirements. First, it has to greatly reduce
the calculational cost. Second, it has to keep essential physics of
transient dynamics. Third, it must have the correct initial current
at $t=0$ and approach the correct asymptotic limit at $t\rightarrow
\infty$. The first goal is realized by eliminating double energy
integral using a reasonable ansatz, with which the dynamical
properties of molecular device is maintained.

To find such an ansatz, we first assume that ${\bf \Sigma}^{a,0}(E)$
changes smoothly and slightly with $E$ and is analytic in the upper
half plane, so that the typical integral like $\int d\epsilon dE~
\frac{e^{i(\epsilon-E)t}}{-i(E-\epsilon+i0^+)}{\bf \Sigma}^{a,0}(E)$
is roughly zero due to the different phase in $e^{\i(\epsilon-E)t}$.
Then the last term of ${\bf F}^{U/D}$ and the second term of ${\bf
Q}^{U/D}$ disappear. Considering the following identity,
\begin{eqnarray}
&&\int \frac{dE}{2\pi}~\frac{{\bf
\Sigma}^a_\alpha(E)}{-i(E-\epsilon+i0^+)}
\nonumber \\
&=&\left[\int_{-\infty}^{0^-}+\frac{1}{2}\int^{0^+}_{0^-}\right]d\tau~
{\bf \Sigma}^a_\alpha(\tau)\int\frac{dE}{2\pi}~
\frac{e^{iE\tau}}{-i(E-\epsilon+i0^+)}
\nonumber \\
&=&\left[\int_{-\infty}^{0^+}-\frac{1}{2}\int^{0^+}_{0^-}\right]d\tau~
e^{i\epsilon \tau} {\bf \Sigma}^a_\alpha(\tau)={\bf
\Sigma}^a_\alpha(\epsilon)-{\bf \Delta}^a_\alpha\nonumber
\end{eqnarray}
and defining ${\bf \Sigma}^a_\alpha(E,\Delta)={\bf
\Sigma}^a_\alpha(E)-{\bf \Delta}^a_\alpha$, the first term of ${\bf
F}_{U/D}$ and ${\bf Q}_{U/D}$ in Eqs.(\ref{expc}) can be simplified,
${\bf F}_{U/D}$ now becomes
\begin{eqnarray}
{\bf F}^D_{\beta\alpha}&\simeq&{\bf G}^{a,0}(\epsilon){\bf
\Sigma}^{a,0}_\alpha(\epsilon,\Delta)+\int\frac{dE}{2\pi}~e^{-i(\epsilon-E)t}
\nonumber\\
&\times&\left[Z^*(\epsilon_\beta)-Z^*(\epsilon)+ {\bf
G}^{a,V}(\epsilon_\beta){\bf P}^\dagger_D \right]{\bf
G}^{a,0}(E){\bf \Sigma}^{a,0}_\alpha(E,\Delta) \label{newFD} \\
{\bf F}^U_{\beta\alpha}&\simeq&{\bf G}^{a,V}(\epsilon_\beta){\bf
\Sigma}^{a,0}_\alpha(\epsilon_{\beta\alpha},\Delta)
+\int\frac{dE}{2\pi}~e^{-i(\epsilon_\beta-E)t}
\nonumber\\
&\times&\left[Z^*(\epsilon)-Z^*(\epsilon_\beta)+ {\bf
G}^{a,0}(\epsilon){\bf P}^\dagger_U\right]{\bf G}^{a,V}(E){\bf
\Sigma}^{a,0}_\alpha(E-eV_\alpha,\Delta) \nonumber\\\label{newFU}
\end{eqnarray}
We note that, in the wide-band limit, Eq.(\ref{newFD},\ref{newFU})
is exact. With our approximation we have eliminated one of the
energy integrals in $J^{out}$, and ${\bf A}$ and ${\bf F}$ now have
similar structures since $\tilde{\bf F}\sim{\bf A}^\dagger {\bf
\Sigma}^a$.

With the approximation defined in Eq.(\ref{newFD},\ref{newFU}), the
current can be written in a compact form (see section C) if we
introduce the effective Green's function
\begin{eqnarray}
&&\tilde{\bf G}^{r/a,0}(E,\epsilon)=\left[E{\bf S}-{\bf
H}^0_c-\sum_\alpha{\bf
\Sigma}^{r/a,0}_\alpha(\epsilon)\right]^{-1}\label{GEep0} \\
&&\tilde{\bf G}^{r/a,V}(E,\epsilon)=\left[E{\bf S}-{\bf
H}^V_c-\sum_\alpha{\bf
\Sigma}^{r/a,V}_\alpha(\epsilon)\right]^{-1}\label{GEepV}
\end{eqnarray}
In general we have to consider the overlap matrix ${\bf S}$.
However, we should keep in mind that in the deriving of the time
dependent current, we have to orthogonalize the basis set, which
would lead to ${\bf S}={\bf I}$. Here $\tilde{\bf
G}^{r/a}(E,\epsilon)$ can be regarded as the Green's functions at
energy $E$ and constant parameter $\epsilon$ for open system with
the effective Hamiltonian
${\bf H}^{r/a}_{eff}={\bf H}_c+{\bf \Sigma}^{r}_\alpha(\epsilon)$.
For a given ${\bf H}_{eff}$, Eqs.(\ref{GEep0},\ref{GEepV}) are
equivalent to
\begin{eqnarray}
(E{\bf S}-{\bf H}^r_{eff})\tilde{\bf G}^r={\bf I}\label{Green2}
\end{eqnarray}
On the other hand, Green's function can be expanded in terms of the
eigenfunctions of the corresponding Hamiltonian,
\begin{equation}
\tilde{\bf G}^r=\sum_n{\bf \Psi}^n C_n. \label{Green1}
\end{equation}
where ${\bf H}_{eff}{\bf \Psi}^n=E_n(\epsilon){\bf \Psi}_n$.
Substituting Eq.(\ref{Green1}) into Eq.(\ref{Green2}), and using the
general orthogonality relation ${\bf \Phi}^{n,\dagger}{\bf S}{\bf
\Psi}^m=C_m\delta_{nm}$ [see Appendix {\ref{Orrelation}] and the
eigenvalue equation ${\bf H}_{eff}{\bf \Psi}^n=E_n(\epsilon){\bf
\Psi}^n$, we have
\begin{eqnarray}
\tilde{\bf G}^r(E,\epsilon)=\sum_n\frac{{\bf \Psi}^n{\bf
\Phi}^{n,\dagger}} {[E-E_n(\epsilon)]{\bf \Phi}^{n,\dagger}{\bf
S}{\bf \Psi}^n}\label{res}
\end{eqnarray}
Obviously, this Green's function can be calculated by finding the
residues ${\rm Res}_n={\bf \Psi}^n{\bf \Phi}^{n,\dagger}/ {{\bf
\Phi}^{n,\dagger}{\bf S}{\bf \Psi}^n}$ at various poles
$E=E_n(\epsilon)$.

Then, we replace $Z(\epsilon){\bf G}^{r/a}(E)$ in
Eqs.(\ref{AD1},\ref{AU1},\ref{newFD},\ref{newFU}) by
$Z(\epsilon)\tilde{\bf G}^{r/a}(E,\epsilon)$. Although $\tilde{\bf
G}^{r/a}(E,\epsilon)$ is different from initial Green's function
${\bf G}^{r/a}(E)=\left[E-{\bf H}_c-{\bf
\Sigma}^{r/a}(E)\right]^{-1}$, this substitution is reasonable since
the major contribution of the integration in
Eqs.(\ref{AD1}-\ref{FU1}) comes from the pole $\epsilon$ in
$Z(\epsilon)$ (see Eq.(\ref{zz})). Similarly, considering the major
contribution of the pole of $Z(\epsilon)$, we replace
$Z(\epsilon){\bf \Sigma}^{a,0}(E)$ in
Eqs.(\ref{AD1},\ref{AU1},\ref{newFD},\ref{newFU}) by
$Z(\epsilon){\bf \Sigma}^{a,0}(\epsilon)$. Since ${\bf
\Sigma}(\epsilon)$ in $\tilde{\bf G}^r(E,\epsilon)$ is independent
of energy $E$, we can perform contour integration over energy $E$ in
Eqs.(\ref{AD1}) and (\ref{AU1}) by closing a contour on lower half
plane and perform the integration over energy $E$ in Eqs.(\ref{FD1})
and (\ref{FU1}) by closing a contour on upper half plane. Thus,
energy integration over $E$ can be analytically performed. It should
be noted that the self energy ${\bf \Sigma}^{r/a}$ is not
independent of energy in contrast to the wide-band limit, this
energy dependence is on $\epsilon$ but not on $E$. In this way, we
can reduce the computational cost and keep the essential physics of
the dynamics as we will show later.

\subsection{Approximate expression of ${\bf A}_\beta(t,\epsilon)$ and ${\bf
F}_{\beta\alpha}(t,\epsilon)$}

Now, considering the initial current and the asymptotic long time
limit, we can write the approximate expression of ${\bf
A}_\beta(t,\epsilon)$ and ${\bf F}_{\beta\alpha}(t,\epsilon)$ from
Eqs.(\ref{AD1},\ref{AU1},\ref{newFD},\ref{newFU}):
\begin{eqnarray}
&&{\bf A}^{D/U}_\beta(t,\epsilon)={\bf
A}^{D/U}_{\beta,1}+{\bf A}^{D/U}_{\beta,2}\label{AF1_A} \\
&&{\bf F}^D_{\beta\alpha}(t,\epsilon)= {\bf
A}^{D,\dagger}_{\beta,1}{\bf
\Sigma}^{a,0}_\alpha(\epsilon_{\beta\alpha},\Delta)+{\bf
A}^{D,\dagger}_{\beta,2}{\bf \Sigma}^{a,0}_\alpha(\epsilon,\Delta)
\label{AF1_FD} \\
&&{\bf F}^U_{\beta\alpha}(t,\epsilon)= {\bf
A}^{U,\dagger}_{\beta,1}{\bf
\Sigma}^{a,0}_\alpha(\epsilon,\Delta)+{\bf
A}^{U,\dagger}_{\beta,2}{\bf
\Sigma}^{a,0}_\alpha(\epsilon_{\beta\alpha},\Delta)\label{AF1_FU}
\end{eqnarray}
with
\begin{eqnarray}
&&{\bf A}^D_{\beta,1}=\int\frac{dE}{2\pi}~
e^{i(\epsilon-E)t}\left[Z(\epsilon_\beta)\tilde{\bf
G}^{r,0}(E,\epsilon_\beta)\left({\bf
I}+{\bf \Xi}^D{\bf G}^{r,V}(\epsilon_\beta)\right)\right] 
\label{AD21}\\
&&{\bf A}^D_{\beta,2}={\bf
G}^{r,0}(\epsilon)-\int\frac{dE}{2\pi}~
e^{i(\epsilon-E)t}\left[Z(\epsilon)\tilde{\bf G}^{r,0}(E,\epsilon)\right]\label{AD22} \\
&&{\bf A}^U_{\beta,1}=\int\frac{dE}{2\pi}~
e^{i(\epsilon_\beta-E)t}\left[Z(\epsilon)\tilde{\bf
G}^{r,V}(E,\epsilon)\left({\bf I}+{\bf \Xi}^U{\bf
G}^{r,0}(\epsilon)\right)\right] 
\label{AU21} \\
&&{\bf A}^U_{\beta,2}={\bf
G}^{r,V}(\epsilon_\beta)-\int\frac{dE}{2\pi}~
e^{i(\epsilon_\beta-E)t}\left[Z(\epsilon_\beta)\tilde{\bf
G}^{r,V}(E,\epsilon_\beta)\right]\label{AU22}
\end{eqnarray}
where
\begin{eqnarray}
{\bf \Xi}^D&=&{\bf U} +\sum_\delta \left[{\bf
\Sigma}^{r,0}_\delta(\epsilon_{\beta\delta}) -{\bf
\Sigma}^{r,0}_\delta(\epsilon_{\beta})\right]\nonumber\\
&=& {\bf U} +\sum_\delta \left[{\bf
\Sigma}^{r,V}_\delta(\epsilon_{\beta}) -{\bf
\Sigma}^{r,0}_\delta(\epsilon_{\beta})\right]\nonumber \\
 {\bf \Xi}^U&=&-{\bf U}
+\sum_\delta \left[{\bf \Sigma}^{r,0}_\delta(\epsilon) -{\bf
\Sigma}^{r,0}_\delta(\epsilon-eV_\delta)\right]\nonumber\\
&=&-{\bf U} +\sum_\delta \left[{\bf \Sigma}^{r,0}_\delta(\epsilon)
-{\bf \Sigma}^{r,V}_\delta(\epsilon)\right]
\end{eqnarray}
This is the second level of approximation. As we will see later that
it is better than the first level approximation described below. Now
we can make further approximation (the first level). To do this, we
note that the Green's function ${\bf G}^{r}$ can be obtained using
the Dyson equation,
\begin{equation}
{\bf G}^{r,tot}={\bf G}^{r,ex}+{\bf G}^{r,ex}{\bf \Xi}{\bf
G}^{r,tot}
\end{equation}
where ${\bf G}^{r,tot}$ is the Green's function of system denoted by
${\bf H}^{tot}={\bf H}^{ex}+{\bf H}'$, ${\bf G}^{r,ex}$ is the
unperturbed Green's function corresponding to ${\bf H}^{ex}$ that
can be exactly solved, ${\bf \Xi}$ is the effective self energy
describing ${\bf H}'$. If we set ${\bf H}^{ex}$ and ${\bf H}^{tot}$
as zero biased open system and $V_\alpha$ biased open system
respectively, we have
\begin{equation}
{\bf G}^{r,tot}={\bf G}^{r,V}(\epsilon)={\bf G}^{r,0}(\epsilon)+{\bf
G}^{r,0}(\epsilon){\bf \Xi}^D{\bf G}^{r,V}(\epsilon)\label{Dyson1}
\end{equation}
Similarly, if we treat ${\bf H}^{ex}$ and ${\bf H}^{tot}$ as
$V_\alpha$ biased open system and zero biased open system,
respectively, we obtain another Dyson equation
\begin{equation}
{\bf G}^{r,tot}={\bf G}^{r,0}(\epsilon)={\bf G}^{r,V}(\epsilon)+{\bf
G}^{r,V}(\epsilon){\bf \Xi}^U{\bf G}^{r,0}(\epsilon)\label{Dyson2}
\end{equation}
Similar to the derivation of the second level of approximation, we
can also replace ${\bf G}^{ex}(\epsilon)$ by $\tilde{\bf
G}^{ex}(E,\epsilon)$ in Eq.(\ref{Dyson1},\ref{Dyson2}) which leads
to
\begin{eqnarray}
&&\tilde{\bf G}^{r,V}(E,\epsilon)\simeq\tilde{\bf
G}^{r,0}(E,\epsilon)\left[{\bf I}+{\bf \Xi}^D{\bf
G}^{r,V}(\epsilon)\right]\nonumber \\
&&\tilde{\bf G}^{r,0}(E,\epsilon)\simeq\tilde{\bf
G}^{r,V}(E,\epsilon)\left[{\bf I}+{\bf \Xi}^U{\bf
G}^{r,0}(\epsilon)\right]\label{Dyson}
\end{eqnarray}
Then, Eqs.(\ref{AD21}) and (\ref{AU21}) can be further approximated
as
\begin{eqnarray}
&&{\bf A}^D_{\beta,1}=\int\frac{dE}{2\pi}~
e^{i(\epsilon-E)t}\left[Z(\epsilon_\beta)\tilde{\bf
G}^{r,V}(E,\epsilon_\beta)\right]\label{AD23} \\
&&{\bf A}^U_{\beta,1}=\int\frac{dE}{2\pi}~
e^{i(\epsilon_\beta-E)t}\left[Z(\epsilon)\tilde{\bf
G}^{r,0}(E,\epsilon)\right] \label{AU23}
\end{eqnarray}
This is the first level of approximation. It is easy to confirm that
when the self-energy is energy independent these two approximations
lead to exactly the same expression of transient current in the
wide-band limit. In the next section we will numerically compare
these two approximations with the exact solution.

\subsection{initial and asymptotic currents}
\label{limit}

We now show that the currents calculated from
Eqs.(\ref{Jin},\ref{Jout},\ref{AF1_A}-\ref{AU22}) and from
Eqs.(\ref{Jin},\ref{Jout},\ref{AF1_A}-\ref{AF1_FU},\ref{AD22},\ref{AU22},\ref{AD23},\ref{AU23})
satisfy the correct current limit at initial $t=0$ and asymptotic
limit $t\rightarrow \infty$ times. Note that the initial current and
asymptotic currents can be calculated from a standard DC transport
nonequilibrium Green's function analysis. It is expected that the
asymptotic current for the downward pulse $J^D_\alpha(t\rightarrow
\infty)$ and initial current for the upward pulse $J^U_\alpha(t=0)$
are zero since there is no bias in the system. Now we discuss the
limiting cases for two versions of approximations developed in
section IIIC.

When $t=0$, $e^{i(\epsilon-E)t}=1$, we can perform integration over
energy $E$ in Eqs.(\ref{AD21}-\ref{AU22}) by closing a contour at
upper half plane, where only a single residual exists at an energy
pole of $Z$. At $t=0$, $\tilde{\bf G}^{r/a,0/V}(E,\epsilon)={\bf
G}^{r/a,0/V}(\epsilon)$, therefore Eqs.(\ref{AD21},\ref{AU21}) and
Eqs.(\ref{AD23},\ref{AU23}) are equivalent. Now we focus on the
current obtained from
Eqs.(\ref{Jin},\ref{Jout},\ref{AF1_A}-\ref{AF1_FU},\ref{AD22},\ref{AU22},\ref{AD23},\ref{AU23}).
After integrating over $\epsilon$, the two terms in
Eqs.(\ref{AD22},\ref{AU22}) cancels to each other, then from
Eq.(\ref{AD23}, \ref{AU23}), ${\bf A}^{D/U}_\beta(t=0)$ becomes
\begin{eqnarray}
{\bf A}^D_\beta(t=0)&=&\tilde{\bf
G}^{r,V}(\epsilon_\beta,\epsilon_\beta)={\bf
G}^{r,V}(\epsilon_\beta)\label{AD3}\\
{\bf A}^U_\beta(t=0)&=&\tilde{\bf G}^{r,0}(\epsilon,\epsilon)={\bf
G}^{r,0}(\epsilon)\label{AU3}
\end{eqnarray}
For ${\bf F}_{\beta\alpha}$, we can perform integration over energy
$E$ by closing a contour at lower half plane. Similarly, there also
exists only a single residual on energy pole $E_Z$ of $Z^*$ in the
lower half plane, and
\begin{eqnarray}
{\bf F}^D_{\beta\alpha}(t=0)&=&\tilde{\bf
G}^{a,V}(\epsilon_\beta,\epsilon_\beta){\bf
\Sigma}^{a,0}_\alpha(\epsilon_{\beta\alpha},\Delta)={\bf
G}^{a,V}(\epsilon_\beta){\bf
\Sigma}^{a,V}_\alpha(\epsilon_\beta,\Delta)\nonumber \\
\label{FD3}\\
{\bf F}^U_{\beta\alpha}(t=0)&=&\tilde{\bf
G}^{a,V}(\epsilon,\epsilon){\bf
\Sigma}^{a,0}_\alpha(\epsilon,\Delta)={\bf G}^{a,0}(\epsilon){\bf
\Sigma}^{a,0}_\alpha(\epsilon,\Delta)\label{FU3}
\end{eqnarray}
Substituting Eq.(\ref{AD3}-\ref{FU3}) into
Eq.(\ref{Jin},\ref{Jout}), and considering
\begin{eqnarray}
&&{\bf \Sigma}^{\gamma,0}_\beta(\epsilon)={\bf
\Sigma}^{\gamma,V}_\beta(\epsilon_\beta)\nonumber\\
&&{\bf G}^{<,0/V}(\epsilon)={\bf
G}^{r,0/V}(\epsilon)\left[\sum_\beta{\bf
\Sigma}^{<,0/V}_\beta(\epsilon)\right]{\bf G}^{a,0/V}(\epsilon)\nonumber \\
&&{\bf \Sigma}^{<,0}_\beta(\epsilon)=f(\epsilon)\left[{\bf
\Sigma}^{a,0}_\beta(\epsilon)-{\bf
\Sigma}^{r,0}_\beta(\epsilon)\right]
\nonumber \\
&&{\bf \Sigma}^{<,V}_\beta(\epsilon)=f(\epsilon-eV_\beta)\left[{\bf
\Sigma}^{a,V}_\beta(\epsilon)-{\bf
\Sigma}^{r,V}_\beta(\epsilon)\right]
\end{eqnarray}
where $f(\epsilon)$ is Fermi distribution function, we have initial
current at $t=0$
\begin{eqnarray}
&&J^D_\alpha=2e\rm{Re}\int \frac{d\epsilon}{2\pi}~{\bf
G}^{r,V}(\epsilon){\bf \Sigma}_\alpha^{<,V}(\epsilon)+{\bf
G}^{<,V}(\epsilon)
{\bf \Sigma}^{a,V}_\alpha(\epsilon)\label{initialD}\\
&&J^U_\alpha=2e\rm{Re}\int \frac{d\epsilon}{2\pi}~{\bf
G}^{r,0}(\epsilon){\bf \Sigma}_\alpha^{<,0}(\epsilon)+{\bf
G}^{<,0}(\epsilon){\bf
\Sigma}^{a,0}_\alpha(\epsilon)\label{initialU}
\end{eqnarray}
Eqs.(\ref{initialD}) and (\ref{initialU}) are the same as the formal
DC current expression in the case of nonzero bias and zero bias,
respectively. $J^U_\alpha(t=0)$ in Eq.(\ref{initialU}) is exactly
zero since the Fermi distribution in ${\bf \Sigma}^<_\alpha$ and
${\bf G}^<$ are equal for $\alpha=L$ and $\alpha=R$.

When $t\rightarrow \infty$, by virtue of the Riemann-Lebesgue
lemma,\cite{lemma} the Fourier integral over $\epsilon$ vanishes,
i.e., $\int \frac{d\epsilon}{2\pi}~e^{-i\epsilon t}{\bf G}^r{\bf
\Sigma}^r...$ equal to zero at $t\rightarrow \infty$ since there
always exist poles in lower half plane. With this in mind, we have
\begin{eqnarray}
{\bf A}^D_\beta(t\rightarrow\infty,\epsilon)&=&{\bf
G}^{r,0}(\epsilon)\label{AD4}\\
{\bf F}^D_{\beta\alpha}(t\rightarrow\infty,\epsilon)&=&{\bf
G}^{a,0}(\epsilon){\bf
\Sigma}^{a,0}_\alpha(\epsilon,\Delta)\label{FD4}\\
{\bf A}^U_\beta(t\rightarrow\infty,\epsilon)&=&{\bf
G}^{r,V}(\epsilon_\beta)\label{AU4}\\
{\bf F}^U_{\beta\alpha}(t\rightarrow\infty,\epsilon)&=&{\bf
G}^{a,V}(\epsilon_\beta){\bf
\Sigma}^{a,0}_\alpha(\epsilon_{\beta\alpha},\Delta)={\bf
G}^{a,V}(\epsilon_\beta){\bf
\Sigma}^{a,V}_\alpha(\epsilon_\beta,\Delta)\nonumber\\\label{FU4}
\end{eqnarray}
From Eq.(\ref{AD4}-\ref{FU4}) and Eq.(\ref{Jin},\ref{Jout}), we have
the asymptotic current
\begin{eqnarray}
&&J^D_\alpha=2e\rm{Re}\int \frac{d\epsilon}{2\pi}~{\bf
G}^{r,0}(\epsilon){\bf \Sigma}_\alpha^{<,0}(\epsilon)+{\bf
G}^{<,0}(\epsilon){\bf
\Sigma}^{a,0}_\alpha(\epsilon)  \label{asymptoticD}\\
&&J^U_\alpha=2e\rm{Re}\int \frac{d\epsilon}{2\pi}~{\bf
G}^{r,V}(\epsilon){\bf \Sigma}_\alpha^{<,V}(\epsilon)+{\bf
G}^{<,V}(\epsilon) {\bf
\Sigma}^{a,V}_\alpha(\epsilon) \nonumber \\
\label{asymptoticU}
\end{eqnarray}
It is easy to see, Eqs.(\ref{asymptoticD}) and (\ref{asymptoticU})
are the formal DC current expression in the case of zero bias and
nonzero bias, respectively, and $J^D_\alpha(t\rightarrow\infty)$ in
Eq.(\ref{asymptoticD}) is exactly zero.

\section{comparison with the exact result in quantum dot system}
Now we consider a system composed of a single-level quantum dot
connected to external leads with a Lorentzian linewidth. This system
can be solved exactly to give a transient current for pulse-like
bias\cite{Maciejko}. We can obtain transient current using three
methods: (i) the exact current expressed by
Eqs.(\ref{Jin},\ref{Jout}, \ref{AD1}-\ref{FU1}), (ii) the first
level of approximation from
Eqs.(\ref{Jin},\ref{Jout},\ref{AF1_A}-\ref{AF1_FU},\ref{AD22},\ref{AU22},\ref{AD23},\ref{AU23})
and (iii) the second level of approximation from
Eqs.(\ref{Jin},\ref{Jout},\ref{AF1_A}-\ref{AU22}). We will compare
the current obtained from these three methods. The system is
described by the following simple Hamiltonian
\begin{eqnarray}
H=\sum_{k_\alpha}\epsilon_{k_\alpha}(t)c^\dagger_{k_\alpha}c_{k_\alpha}+
\epsilon_d(t)d^\dagger
d+\sum_{k_\alpha}(t_{k_\alpha}c^\dagger_{k_\alpha}d+h.c.)
\end{eqnarray}
where $\epsilon_d(t)=\epsilon_d^0+U(t)$ and
$\epsilon_{k_\alpha}(t)=\epsilon^0_{k_\alpha}+V_\alpha(t)$. Because
the scattering region has only one state with energy level
$\epsilon^0_d$, the Green's functions $G(\epsilon)$ and self energy
$\Sigma(\epsilon)$ thus become scalars instead of matrices. If we
choose linewidth function
$\Gamma_\alpha(\omega)\equiv2\pi\rho_\alpha(\omega)|t_{k_\alpha}|^2$
to be Lorentzian with the linewidth amplitude $\Gamma^0_\alpha$,
$$\Gamma_\alpha(\omega)=\frac{W^2}{\omega^2+W^2}\Gamma^0_\alpha$$
then $G^\gamma(\epsilon)$ and $\Sigma^\gamma(\epsilon)$ can be
expressed as
\begin{eqnarray}
&&G^{r/a,0}(\epsilon)=\left[\epsilon-\epsilon_d^0-\sum_\alpha\Sigma^{r/a,0}(\epsilon)\right]^{-1}\nonumber \\
&&G^{r/a,V}(\epsilon)=\left[\epsilon-\epsilon_d^0-U^V-\sum_\alpha\Sigma^{r/a,V}(\epsilon)\right]^{-1}\nonumber \\
&&G^{<,0/V}(\epsilon)=
G^{<,0/V}(\epsilon)\left[\sum_\alpha\Sigma^{<,0/V}(\epsilon)\right]G^{<,0/V}(\epsilon)\nonumber \\
&&\Sigma^{r/a,0}_\alpha(\epsilon)=\int
\frac{d\omega}{2\pi}~\Gamma_\alpha(\omega)/(\epsilon-\omega\pm i0^+)\nonumber\\
&&\Sigma^{r/a,V}_\alpha(\epsilon)=\int
\frac{d\omega}{2\pi}~\Gamma_\alpha(\omega)/(\epsilon-eV_\alpha-\omega\pm i0^+)\nonumber\\
&&\Sigma^{<,0}_\alpha(\epsilon)=f(\epsilon)
\left[\Sigma^{a,0}_\alpha(\epsilon)-\Sigma^{r,0}_\alpha(\epsilon)\right]\nonumber\\
&&\Sigma^{<,V}_\alpha(\epsilon)=f(\epsilon-eV_\alpha)
\left[\Sigma^{a,V}_\alpha(\epsilon)-\Sigma^{r,V}_\alpha(\epsilon)\right]\nonumber
\end{eqnarray}
Using the theorem of residual, we can analytically perform integral
in $A_\beta$ and $F_{\beta\alpha}$ for either exact formula or two
approximate formulas. In the calculation, we set
$\Gamma=\Gamma^0_L+\Gamma^0_R$ as the energy unit, and set
$\Gamma^0_L=\Gamma^0_R=0.5$.

We first consider the transient current induced by opposite voltage
$V_L(t)=-V_R(t)$. In this case, the equilibrium coulomb potential in
quantum dot $U^{0,V}=0$, and the time dependent perturbation coming
from coulomb response $U(t)$ is assumed to be zero. It is a
reasonable assumption since the coulomb potential in scattering
region is canceled by the opposite voltage in left and right lead.
In Fig.1, we plot two approximated transient currents and exact
transient current in downward [panel (a), (b), (c)] and upward
[panel (d), (e), (f)] case vs time for different bandwidth $W$. We
find that for all bandwidth $W$, the approximated current and exact
current have the same dynamical behaviors. Fig.2 gives direct
comparison where we merge panels (a), (b) and (c) in Fig.1 as panel
(a) in Fig.2, and merge panels (d), (e) and (f) in Fig.1 as panel
(b) in Fig.2. We can see that for the downward pulse [panel (a)],
transient current using three formulas are almost indistinguishable.
This means that in the opposite voltage, our approximation, the
first approximation
[Eqs.(\ref{AD22},\ref{AU22},\ref{AD23},\ref{AU23})] and the second
approximation [Eqs.(\ref{AD21}-\ref{AU22})] are all very good for
studying transient dynamics. For the upward pulse, although the
approximations are not as good as in downward case, the currents
calculated from approximate scheme are still in good agreement with
the exact solution especially for the second approximation. Hence we
may conclude that the two approximations are all reasonable in the
opposite voltage $V_L(t)=-V_R(t)$. They can be used to study
transient dynamics in the real molecular device to speed up the
calculation.
\begin{figure}
\includegraphics[bb=10mm 10mm 184mm 185mm,
width=8.5cm,totalheight=8cm, clip=]{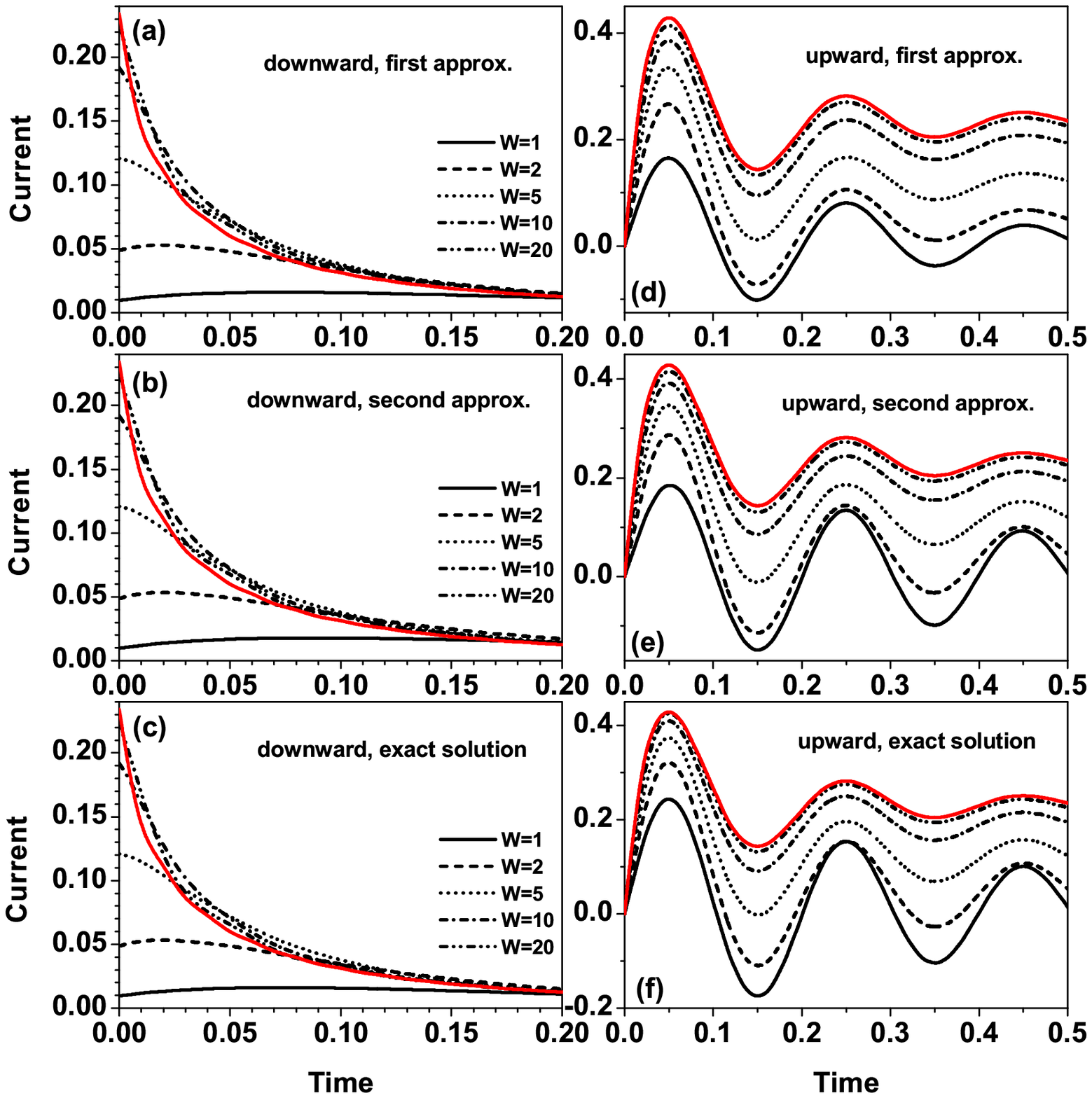} \caption{ (Color
online) Time dependent current $J(t)$ corresponding to an opposite
downward pulse or upward pulse in three versions: the exact solution
and two approximations. The different black lines are for different
bandwidth $W$. The red line is for $W=\infty$, i.e., the wide-band
limit. The current is in the unit of $e\Gamma$, and the time is in
the unit of $2\pi/\Gamma$. $eV_L=-eV_R=5.$}
\end{figure}
\begin{figure}
\includegraphics[bb=11mm 10mm 213mm 255mm,
width=6.5cm,totalheight=8.5cm, clip=]{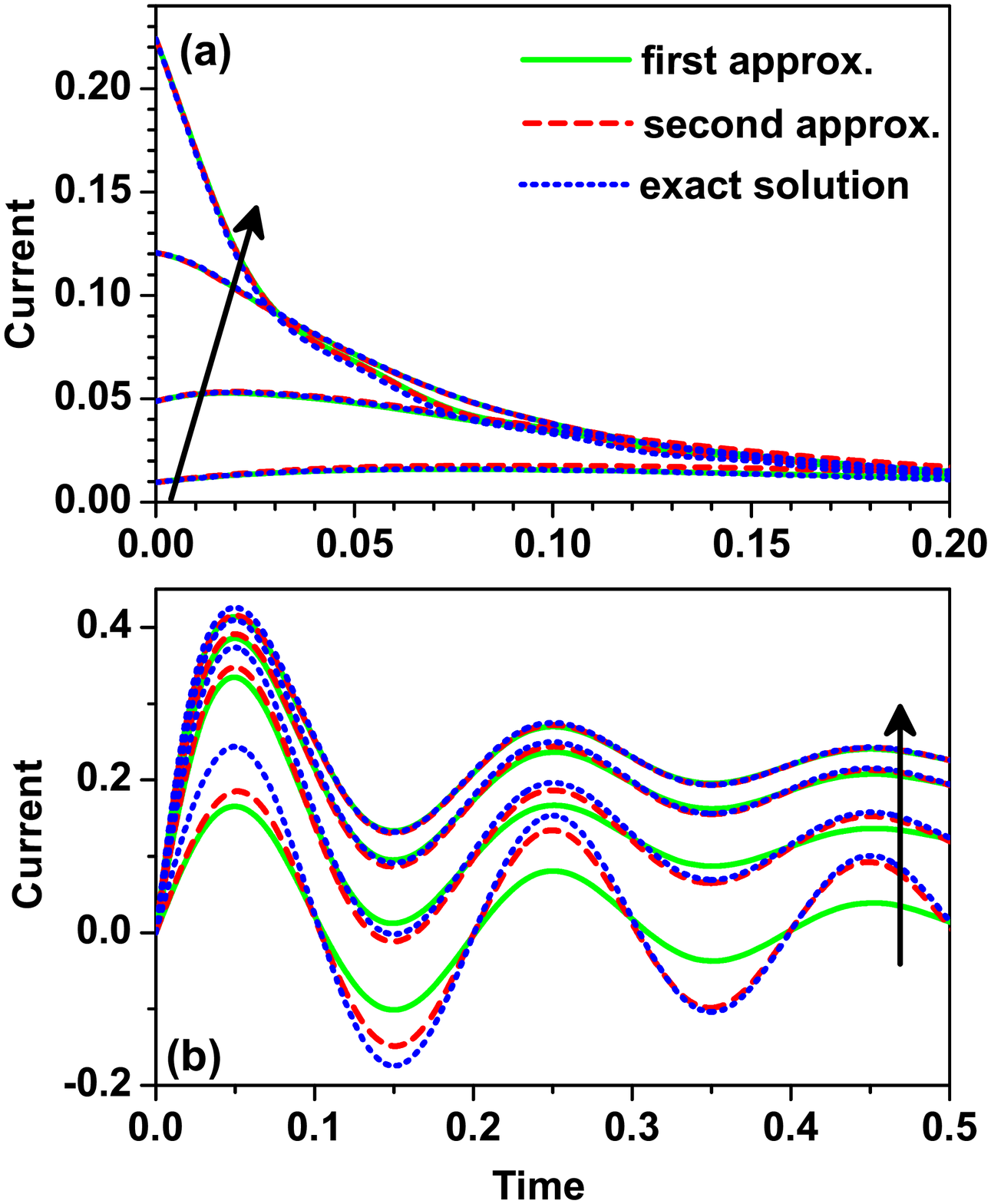} \caption{(Color
online) Merged version of Fig.1 for $W=1$, $2$, $5$ and $20$. Panel
(a) corresponding to the downward pulse current comes from panel
(a), (b) and (c) in Fig.1, panel (b) corresponding to upward pulse
current comes from panel (d), (e) and (f). Along the black arrow,
the bandwidth are $W=1$, $2$, $5$ and $20$, respectively.}
\end{figure}

Next, we focus on the asymmetric voltage, i.e., $V_L(t)\neq V_R(t)$.
In this case, the equilibrium coulomb potential in quantum dot
$U^{0/V}$, and the time dependent perturbation coming from coulomb
response $U(t)$ can't be canceled by the voltage in left and right
lead. In principle, perturbation $U(t)$ should be calculated by
solving time dependent Schr\"{o}dinger equation, it will be very
difficult and computational demanding therefore can't be implemented
in real molecular device. As an alternative scheme, we have set
$U(t)=[eV_L(t)\Gamma^0_L +eV_L(t)\Gamma^0_L ]/\Gamma$. For the
single level quantum dot system, this is exact because the central
scattering region now is expressed in a scalar instead of matrices,
which leads to the same transient current for the opposite voltage
$V_L(t)=-V_R(t)$ and asymmetric voltage $V_L(t)=V(t)$, $V_R(t)=0$ or
$V_L(t)=0$, $V_R(t)=-V(t)$ in the exact solution.

For the first approximation the poles in time dependent term
$e^{i(\epsilon-E)t}$ are different from that in the second level
approximation, i.e., the poles of $\tilde{\bf G}^{r,0}$ in
Eq.\ref{AD21} and $\tilde{\bf G}^{r,V}$ in Eq.\ref{AU21} are
replaced by the poles of $\tilde{\bf G}^{r,V}$ in Eq.\ref{AD23} and
$\tilde{\bf G}^{r,0}$ in Eq.\ref{AU23}, respectively. Because of
this, the time evolution process are not as accurate in the first
approximation, especially for the large $V_\alpha$. So, for the
asymmetric voltage, the second approximation is better. In Fig.3 and
Fig.4, we compare the transient current obtained from the second
approximation [panel (b-d)] for opposite or asymmetric voltage with
the exact transient current [panel (a)] in response to the downward
pulse and upward pulse, respectively. We find that all transient
currents from the second approximation in Fig.3 and Fig.4 [panel
(b)] are very close to the exact result [panel (a)]. Moreover, in
Fig.3 and Fig.4, the approximate transient current in panel (b),
(c), (d) have almost the same behavior. It is safe to say that our
approximations have kept essential physics of dynamical transport
properties.

\begin{figure}
\includegraphics[bb=11mm 10mm 186mm 136mm,
width=8.5cm,totalheight=6.2cm, clip=]{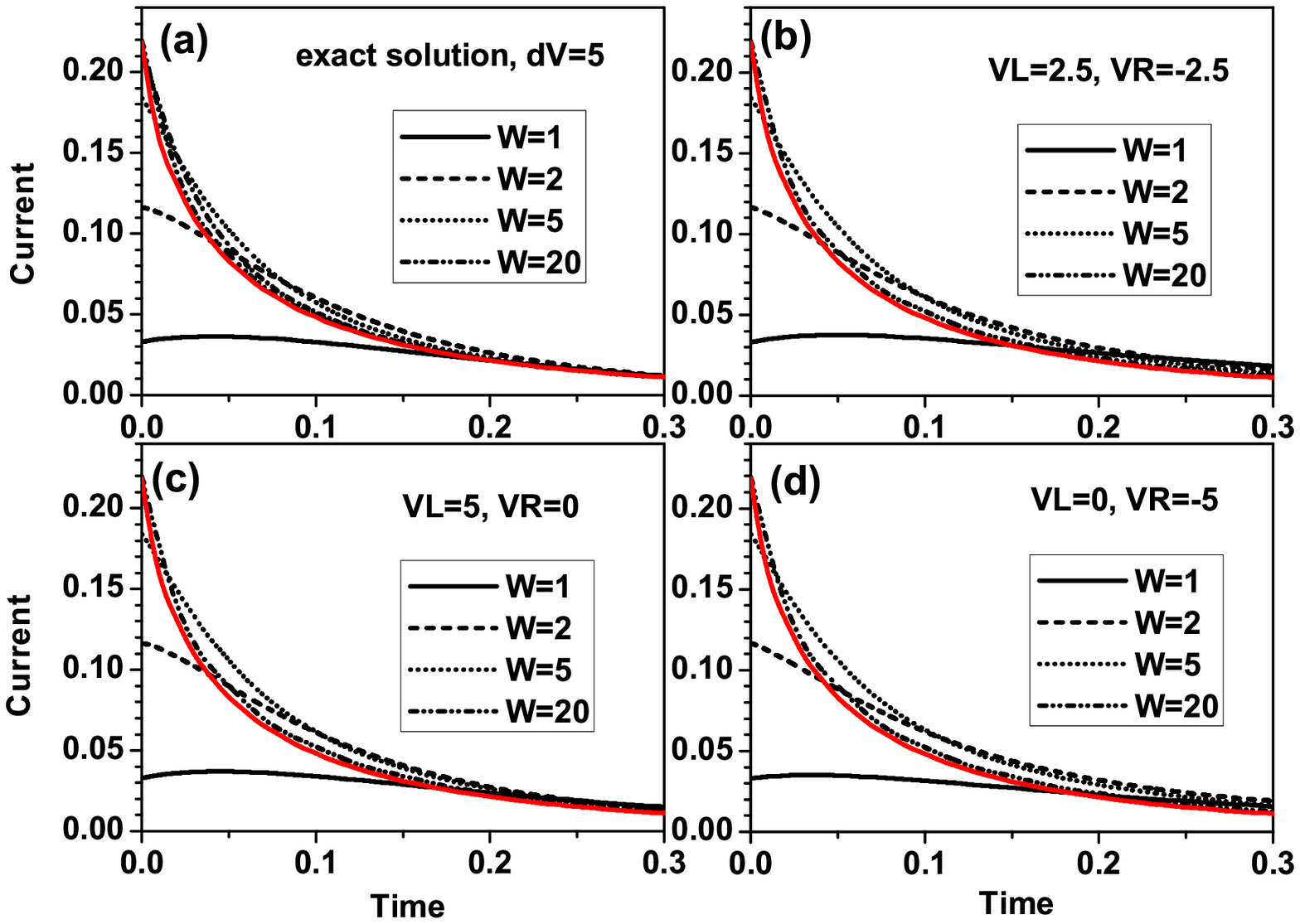} \caption{(Color
online) Panel (a): exact time dependent current $J(t)$ corresponding
to downward pulse for $dV=V_L-V_R=5$. Panel (b-d) are corresponding
to the second approximate transient current corresponding to
downward pulse for opposite voltage $V_L=-V_R=2.5$, asymmetric
voltage $V_L=5$, $V_R=0$ and $V_L=0$, $V_R=-5$, respectively. The
different black lines are for different bandwidths $W$. The red line
is wide-band limit for $W=\infty$.} \label{Fig3_1}
\end{figure}
\begin{figure}
\includegraphics[bb=12mm 11mm 183mm 136mm,
width=8.5cm,totalheight=6.2cm, clip=]{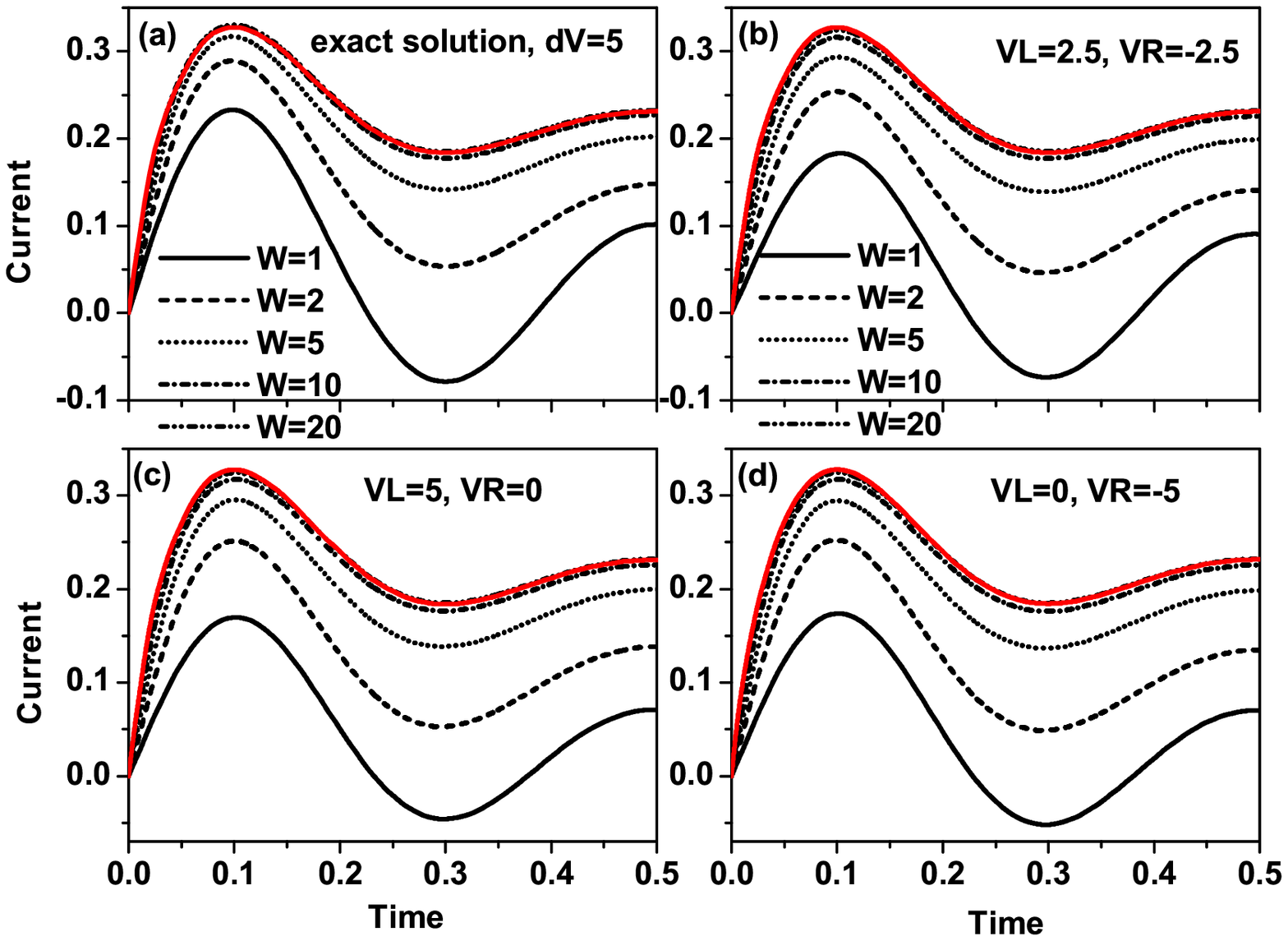} \caption{(Color
online) Same to Fig.\ref{Fig3_1}, transient current corresponding to
upward pulse vs time are plotted.}
\end{figure}

\section{several examples for real molecular devices}

In this section, we implement our approximate formula in two
representative molecular devices including a short carbon chain
coupled to aluminum leads and a $C_{60}$ molecule coupled to
aluminum leads. These systems were chosen because they are typical
in first-principles calculation and their practical importance to
nano-electronics. In Fig.5(a) and Fig.5(b), we show the structure of
Al-${\rm C}_5$-Al and Al-${\rm C}_{60}$-Al, respectively, where Al
leads are along (100) direction, one unit cell of Al lead consists
of 9 Al atoms and total 40 atoms were included in the simulation
box. For the Al-C5-Al device, the nearest distance between Al leads
and the carbon chain is 3.781 a.u. and the distance of C-C bond is
2.5 a.u.(1 a.u.=0.529{\AA}). In Al-C60-Al device, the distance
between the Al atom and the nearest C atom equal to 3.625 a.u..

To calculate the dynamic response of molecular devices, we have used
the first-principles quantum transport package
MATDCAL.\cite{matdcal} Considering the complicated coulomb response
in scattering region, we set $V_L(t)=-V_R(t)$. In this case, the
first approximation is simple but as good as the second one. So, in
the following, the first approximate formula
[Eqs.(\ref{Jin},\ref{Jout},\ref{AF1_A}-\ref{AF1_FU},\ref{AD22},\ref{AU22},\ref{AD23},\ref{AU23})]
is used. In principle, the calculation involves the following steps:
(1) calculate the device Hamiltonian including central scattering
Hamiltonian and lead Hamiltonian using NEGF-DFT package to get two
potential landscapes $U^0$ at zero bias and $U^V$ at $V_\alpha$
bias, respectively. They are originally expressed in a nonorthogonal
fireball basis. (2) orthogonalize the nonorthogonal device
Hamiltonian using the approach\cite{thesis} introduced in Appendix
\ref{Orbasis} so that they are finally expressed in an orthogonal
basis. (3) with the orthogonal lead Hamiltonian $H_\alpha$, one
calculates zero biased self energy ${\bf \Sigma}^{r/a,0}_\alpha$ and
$V_\alpha$ biased self energy ${\bf \Sigma}^{r/a,V}_\alpha$ from
Eqs.(\ref{dcself},\ref{surfself}) using the transfer matrix
method.\cite{transfer} (4) with orthogonalized central scattering
Hamiltonian ${\bf H}^0_c$ and ${\bf H}^V_c$ and self energy ${\bf
\Sigma}^{r/a,0}_\alpha$ and ${\bf \Sigma}^{r/a,V}_\alpha$ obtained
from two potential landscapes $U^0$ and $U^V$, one solves the
effective Green's function ${\bf G}^{r/a,0/V}$ using
Eqs.(\ref{GEep0},\ref{GEepV}) by calculating its poles and residuals
from Eq.(\ref{res}). Step (1)-(4) are time independent processes and
easy to perform. (5) calculate time dependent quantities ${\bf
A}^{D/U}_{\beta,1}$ and ${\bf A}^{D/U}_{\beta,2}$ from
Eqs.(\ref{AD23},\ref{AU23}) and Eqs.(\ref{AD22},\ref{AU22}). Then
${\bf A}_{\beta}$ and ${\bf F}_{\beta\alpha}$ can be calculated from
Eqs.(\ref{AF1_A}-\ref{AF1_FU}). (6) integrate over $\epsilon$ and
obtain the final AC current
$J^{D/U}(t)=[J^{D/U}_L(t)-J^{D/U}_R(t)]/2$ from
Eqs.(\ref{Jin},\ref{Jout}).

\begin{figure}
\includegraphics[bb=10mm 12mm 188mm 130mm,
width=6.5cm,totalheight=4cm, clip=]{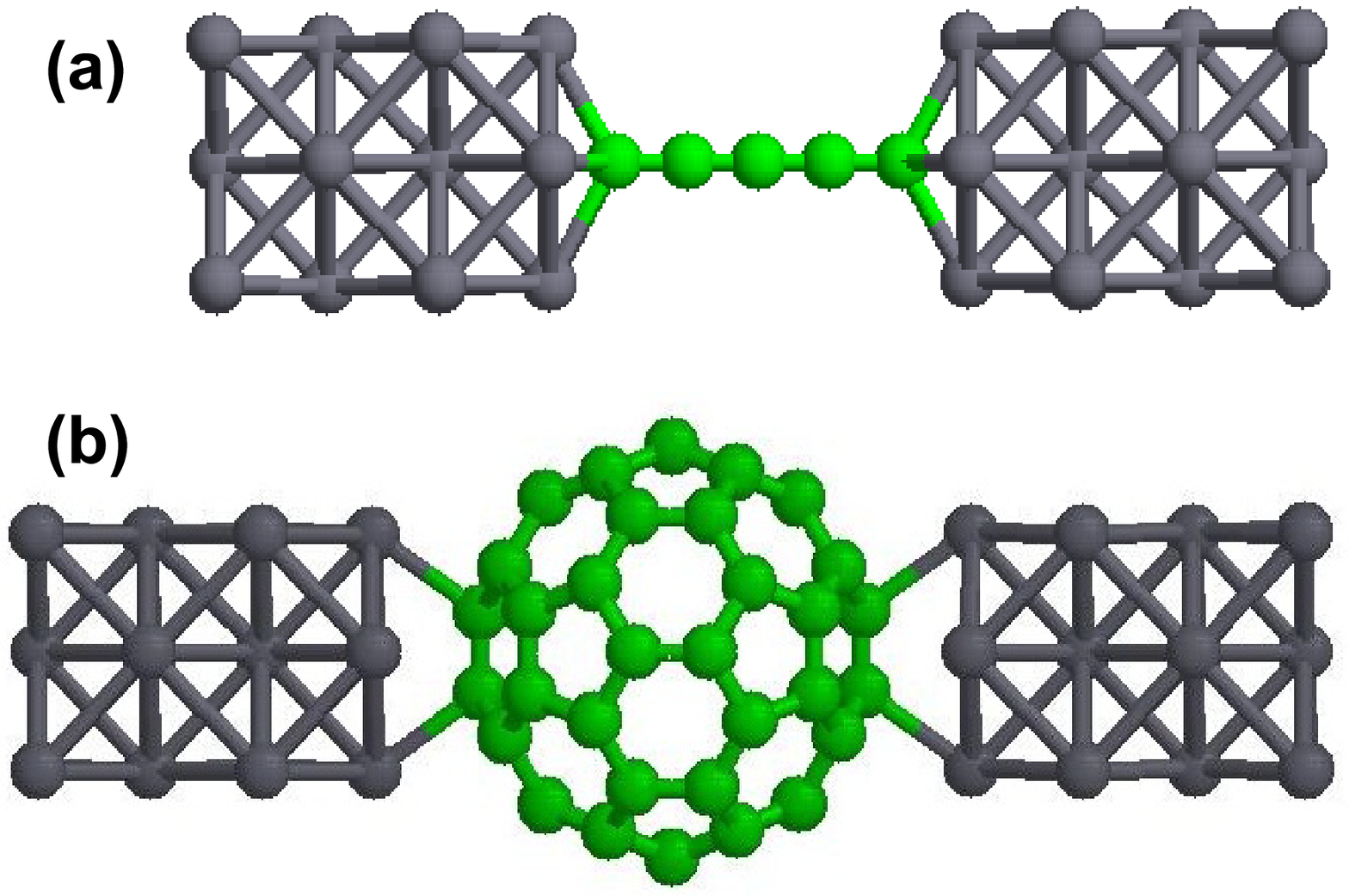} \caption{(Color
online) Panel (a): Structure of Al-C5-Al. Panel(b): structure of
Al-${\rm C}_{60}$-Al.}
\end{figure}

First we study the Al-${\rm C}_5$-Al structure. In Fig.6, we plot
the transient current $J(t)$ corresponding to the upward pulse
[panel (a) and (b)] and the downward pulse [panel (c) and (d)] for
different external voltages $V_R=-V_L=0.001a.u.$ [panel (a) and (c)]
and $V_R=-V_L=0.01a.u.$ [panel (b) and (d)] in Al-${\rm C}_5$-Al
structure. Following observations are in order: (1) as we have
discussed in Sec.\ref{limit}, for all bias voltages $V_\alpha$ the
transient currents indeed reach the correct limits at $t=0$ and
$t\rightarrow\infty$. For the upward pulse, $J(t=0)=0$ and
$J(t\rightarrow\infty)=J_{dc}$ while for the downward pulse we have
$J(t=0)=J_{dc}$ and $J(t\rightarrow\infty)=0$. (2) for both upward
pulse (turn-on voltage) and downward pulse (turn-off voltage), once
the bias voltage is switched, currents oscillate rapidly in the
first a few or tens fs and then gradually approach to the
steady-state values, i.e., $J_{dc}$ for turn-on voltage and zero for
turn-off voltage. The larger the voltage $V_\alpha$, the more rapid
the current oscillates. (3) concerning the long time behavior, the
time dependent current oscillates with a frequency proportional to
$|V_\alpha|$.\cite{WangBin} This is because the time dependent term
$e^{i(\epsilon-E)t}$ in
Eqs.(\ref{AD22},\ref{AU22},\ref{AD23},\ref{AU23}) are $V_\alpha$
dependent. For the upward pulse, $e^{i(\epsilon_\alpha-E)t}\propto
e^{iV_\alpha t}$, which directly leads to the oscillating frequency
proportional to $|V_\alpha|$. For the downward pulse, although
$e^{i(\epsilon-E)t}$ is $V_\alpha$ independent, in the energy
integral on $E$, the pole $E_n$ of $\tilde{\bf G}^r(E,\epsilon)$ are
determined by the self energy ${\bf \Sigma}^{r,V}_\alpha$. Since
${\bf \Sigma}^{r,V}_\alpha$ depends on $V_\alpha$, this leads to
$V_\alpha$ dependent oscillating frequency. In addition, we notice
that although the properties of dc conductance of short carbon
chains are different for the chains with odd and even number
atoms\cite{WangBin2} due to the completely different electronic
structure near Fermi level, the ac signals are similar (see
Ref.\onlinecite{WangBin} where Al-${\rm C}_4$-Al structure was
analyzed). This indicates that in AC transport, all states with
energy from $-\infty$ to the Fermi energy are contributing, which is
very different from dc case where only the states near Fermi level
contribute to transport processes.

\begin{figure}
\includegraphics[bb=11mm 77mm 194mm 211mm,
width=8.5cm,totalheight=6cm, clip=]{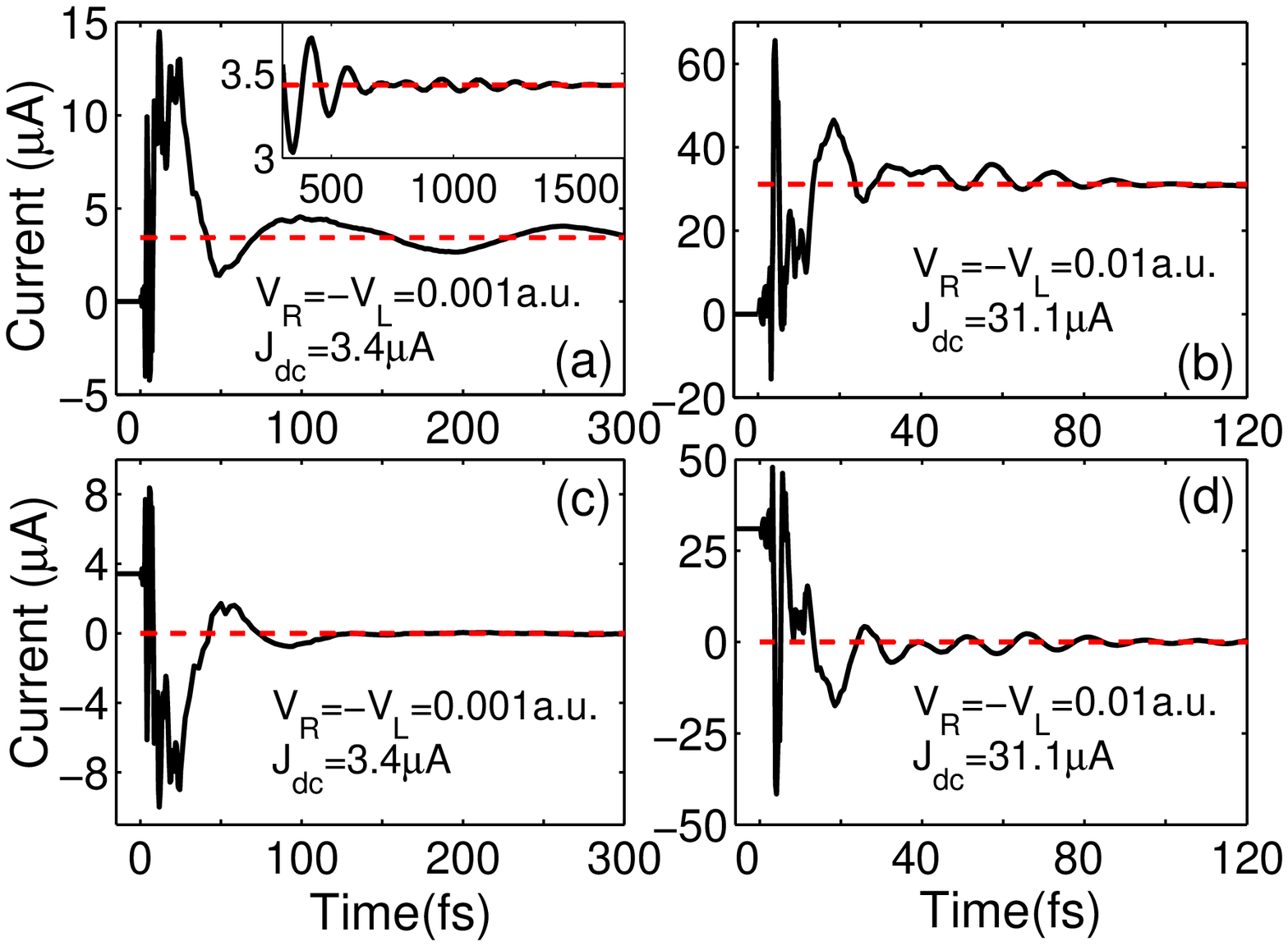} \caption{(Color
online) Time dependent current $J(t)$ corresponding to the upward
pulse [panel (a) and (b)] and the downward pulse [panel (c) and (d)]
for different external voltages $V_\alpha$ in Al-${\rm C}_5$-Al
device. The inset of panel (a) shows the long time behavior of the
time-dependent current. The red (gray in print) dashed lines in
panels indicate asymptotic current $J(t\rightarrow\infty)$ which the
DC current biased by $V_{L/R}$ labeled in corresponding panels for
the upward pulse, and arrive at zero for the downward case.}
\end{figure}

Next, we study the second sample: the Al-${\rm C}_{60}$-Al
structure. In Fig.7, the transient current $J(t)$ of the structure
corresponding to an upward pulse [panel (a) and (b)] and a downward
pulse [panel (c) and (d)] for different external voltages
$V_R=-V_L=0.001a.u.$ [panel (a) and (c)] and $V_R=-V_L=0.01a.u.$
[panel (b) and (d)] are plotted. Similar to the Al-${\rm C}_{5}$-Al
structure, correct initial current $J(t=0)$ and asymptotic current
$J(t\rightarrow\infty)$ are also obtained in Al-${\rm C}_{60}$-Al
structure. In addition, there are also rapidly oscillations at short
times after the switch although the oscillation is not as rapid as
that in the Al-${\rm C}_{5}$-Al structure. Furthermore, similar to
Al-${\rm C}_{5}$-Al structure, in gradually reaching the
steady-state values, the current oscillates with a frequency
proportional to $|V_\alpha|$ but its decay rate is much slower than
that in Al-${\rm C}_{5}$-Al structure. It indicates that there are
much more quasi-resonant state that contribute to the transient
current in Al-${\rm C}_{60}$-Al structure which is reasonable
considering the complex electronic structure of isolated ${\rm
C}_{60}$. In the following,  we will analyze in detail how the
current decays for the Al-${\rm C}_{60}$-Al structure.

\begin{figure}
\includegraphics[bb=12mm 78mm 194mm 210mm,
width=8.5cm,totalheight=6cm, clip=]{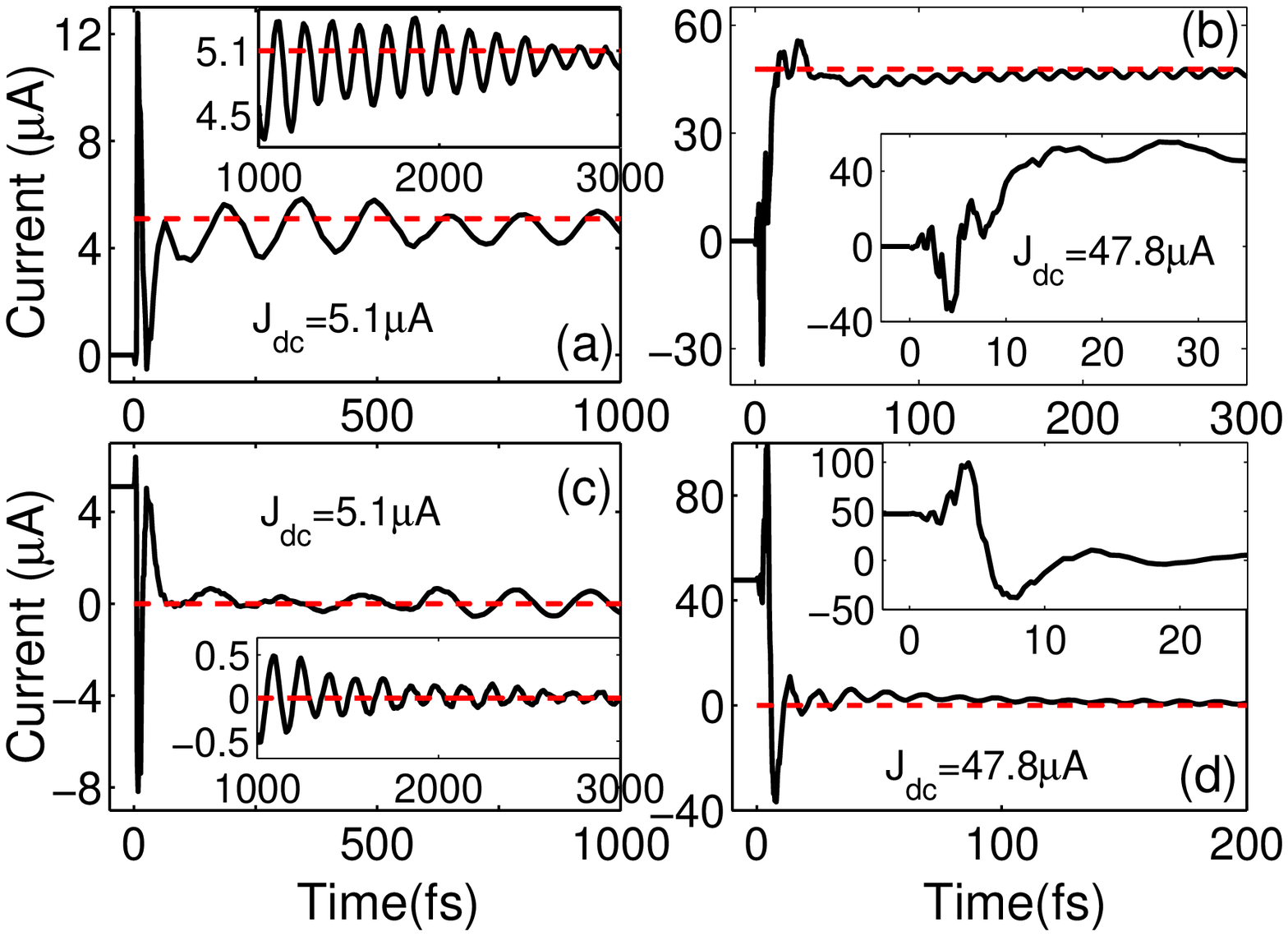} \caption{(Color
online) Time dependent current $J(t)$ corresponding to the upward
pulse [panel (a) and (b)] and the downward pulse [panel (c) and (d)]
in Al-${\rm C}_{60}$-Al device for different $V_\alpha$. In panel
(a) and (c), $V_R=-V_L=0.001a.u.$. In panel (b) and (d),
$V_R=-V_L=0.01a.u.$. Same to Fig.6, the red (gray in print) dashed
lines in panels indicate asymptotic current $J(t\rightarrow\infty)$.
The long time AC current or detailed short time AC current are shown
in inset of panels.}
\end{figure}

Physically, decay time of current corresponds to the width of the
quasi-bound state. In molecular devices, because the linewidth
function ${\bf \Gamma}(\epsilon)$ are complex and energy dependent
matrix, we can't extract characteristic time scale directly from
$1/{\bf \Gamma}$. As such, the transmission coefficient
$T(\epsilon)$ is needed to understand the resonant state and
corresponding characteristic time scale. In Fig.8(a), we plot
transmission coefficient $T(\epsilon)$ in the energy range from the
energy band bottom to the Fermi energy for Al-${\rm C}_{60}$-Al
structure at zero bias. Here, the sharp peaks [some of them, see red
crossed signed peaks in Fig.8(a)] correspond to resonant states with
large lifetimes. At a particular resonant state, the incoming
electron can dwell for a long time, which contributes to a much more
slowly decaying current than other non-resonant states. In Fig.8(b),
(c) and (d), we amplify the first, second and forth labeled
quasi-resonant transmission, respectively, where the peaks' width
$\Gamma_{peak}\sim10^{-5}a.u.$ are indicated, corresponding to a
decay time $\tau\sim 2400 fs$ from the expression
$\Gamma_{peak}t=1$. In Fig.8(e)-(g), corresponding to different
$\epsilon$ where the resonant peaks in Fig.8(b)-(d) are located, we
plot long time behavior of current element $J_L(\epsilon)$. Here
$J_L(\epsilon)$ is the time dependent current for each energy
$\epsilon$, the integration over which gives the final current
$J_\alpha(t)$. We can see that for each resonant state the current
$J_L(\epsilon)$ keeps oscillating in a long time comparable to the
decay time $\tau\sim 2400fs$. Furthermore the intensity of the
oscillation $\Delta J\sim 0.2\mu A$ is not very small comparing to
the DC signal $J_{dc}=5.1\mu A$.

\begin{figure}
\includegraphics[bb=9mm 79mm 193mm 209mm,
width=8.5cm,totalheight=6cm, clip=]{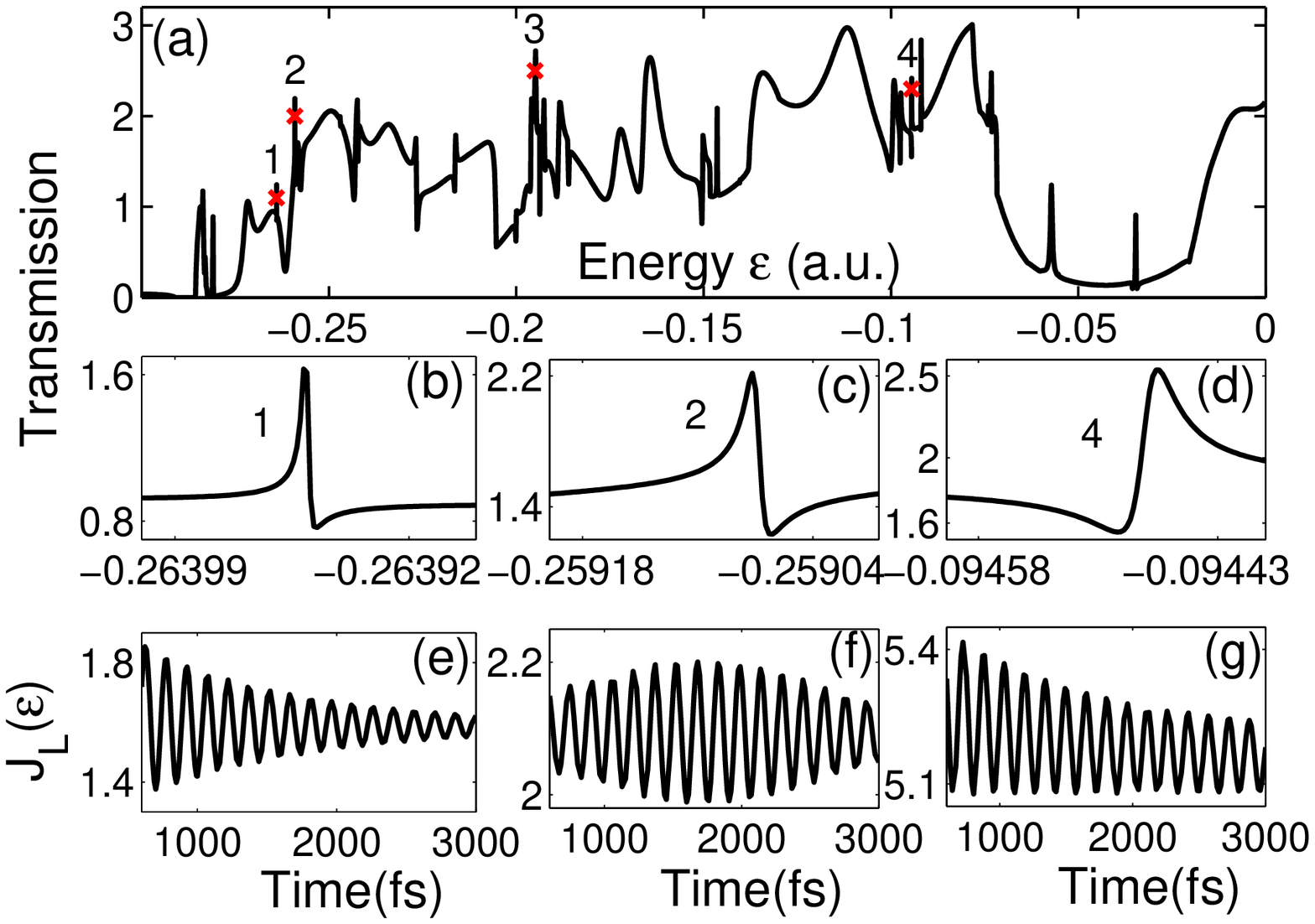} \caption{(Color
online) Panel (a): transmission coefficient $T(\epsilon)$ in the
energy range from the energy band bottom to the Fermi energy. In the
whole energy range, there are some resonant states corresponding to
the very sharp transmission coefficient $T(\epsilon)$, as we have
indicated (see red cross) and labeled (by 1, 2, 3 and 4) in panel
(a), some of them contribute to the current at long time. We amplify
the first, second and forth labeled resonant transmission in panel
(b), (c) and (d), respectively. In panel (e)-(g), we plot the long
time behavior of current $J_L(\epsilon)$ at a fixed $\epsilon$ for
the first, second and forth resonant states. The external voltage
$V_\alpha=0.001a.u.$.}
\end{figure}

After integration over energy, these slowly decaying currents
$J_L(\epsilon)$ due to the resonant states may cancel to each other
partially due to the difference in their phases. However, we should
keep in mind that it is these resonant peaks that may give rise to
convergence problem. Hence in the calculation, we should first scan
the equilibrium and non-equilibrium transmission coefficient
(100,000 energy points for example) to resolve sharp resonant peaks
in the whole energy range from minimum energy to Fermi energy. Then,
for each sharp resonant peak, enough (100 for example) energy points
should be chosen to converge the integration of the current
$J_L(\epsilon)$ over $\epsilon$, i.e., $\int d\epsilon J(\epsilon)$.
For the non-resonant state, i.e., the smoothly changed region in
$T(\epsilon)$, the current $J(\epsilon)$ are integrated using less
energy points.

As we have discussed that the resonant states are important for the
transient current and they must be carefully treated in calculation.
However, in the calculation of the effective Green's function
$\tilde{\bf G}^{r/a,0/V}$, a small imaginary part that is usually
added to the real energy $\epsilon\rightarrow \epsilon+i\eta$ to
help resolving the retarded or advanced self-energies. This in turn
introduces pseudo resonant states. In order to eliminate the pseudo
resonant state in effective Green's function $\tilde{\bf
G}^{r/a,0/V}$ [Eqs.(\ref{GEep0},\ref{GEepV})], one has to calculate
the self-energy by setting $\eta=0$ and resolve the retarded or
advanced self-energies with the aid of the group velocity
$v_k=(\partial E(k)/\partial k)$.\cite{Sanvito}

\section{conclusion}

By orthogonalizing the Hamiltonian expressed in the nonorthogonal
basis and considering the singularity of self-energy
${\Sigma}^{r/a}(t,t')$ at $t=t'$, we have generalized the solution [
developed in Ref.\onlinecite{Maciejko}] of the transient current
driven either by a downward step voltage pulse or by a upward step
pulse. This generalized result can be applied to both the quantum
dot model and real molecular device. Based on the exact solution
given in Ref.\onlinecite{Maciejko}, we derived two approximate
formulas that are suitable for numerical calculation of the
transient current for molecular devices. We have tested our
approximate formula in a quantum dot system where exact numerical
solution exists. For the quantum dot system, we chose a Lorentzian
linewidth (beyond wideband limit) and compared the time-dependent
current calculated using both exact formula and our approximate
formula. We found that for the opposite voltage $V_L(t)=-V_R(t)$,
the results obtained from the exact formalism and two approximate
scheme agree very well with each other especially in the downward
pulse case. For the nonsymmetric voltage $V_L(t)=V(t)$, $V_R(t)=0$
or $V_L(t)=0$, $V_R(t)=-V(t)$, the second approximation is better.
This shows that our approximate formulas captured the essential
physics of the transient current. In addition, it gives the correct
initial current at $t=0$ and correct asymptotic current at
$t\rightarrow \infty$. Since we have reduced the calculation from
triple integral to single integral over the energy, the approximated
approach reduces the computational cost drastically and it can be
easily implemented in first principles calculation for molecular
devices. To demonstrate this, we calculated the transient current
using the first approximated scheme with an opposite voltage
$V_L(t)=-V_R(t)$ for two molecular structures: Al-${\rm C}_{5}$-Al
and Al-${\rm C}_{60}$-Al. Different from the quantum dot system,
because of the complex electronic structure in molecular devices,
transient currents oscillate rapidly in the first a few or tens fs
as the bias voltage is switched, then gradually approach to the
steady-state values. Furthermore, due to the resonant state in
molecular devices, transient currents have a very long decay time
$\tau$.

\begin{appendix}

\section{orthogonality relation for the nonorthogonal basis}
\label{Orrelation} For a system described by $H$, the time
independent eigenvalue equation is written as:
\begin{eqnarray}
H|n\rangle&=&E_n |n\rangle
\end{eqnarray}
the eigenvectors $|n\rangle$ form an orthogonal complete basis set.
However, in many systems such as a molecular device connected to
external leads, the basis set constructed by eigen vectors is not
convenient. We usually expand the eigen vector $|n\rangle$ in other
basis $|\mu\rangle$, which is non-orthogonal complete set (or nearly
complete).
\begin{eqnarray}
|n\rangle\simeq\sum_\mu |\mu\rangle\langle\mu|n\rangle
\end{eqnarray}
the eigenvalue equation now becomes
\begin{eqnarray}
\sum_\mu H|\mu\rangle\langle\mu|n\rangle&=&E_n\sum_\mu
|\mu\rangle\langle\mu|n\rangle\nonumber \\
\sum_\mu \langle\nu |H|\mu\rangle\langle\mu|n\rangle&=&E_n\sum_\mu
\langle\nu|\mu\rangle\langle\mu|n\rangle\nonumber \\
\sum_\mu {\bf H}_{\nu\mu}{\bf \Psi}_\mu^n&=&E_n\sum_\mu {\bf
S}_{\nu\mu}{\bf \Psi}_\mu^n
\end{eqnarray}
where ${\bf S}_{\nu\mu}=\langle\nu|\mu\rangle$. In the matrix form,
we have ${\bf H}{\bf \Psi}^n=E_n{\bf S}{\bf \Psi}^n$. If we use the
self-energy to replace the effect of leads the effective Hamiltonian
for the open system becomes ${\bf H}={\bf H}_0+{\bf \Sigma}^r$.
Since the effective Hamiltonian is not Hermitian, we can define the
adjoint operator ${\bf H}^\dagger={\bf H}={\bf H}_0+{\bf \Sigma}^a$
and corresponding eigen-equation becomes ${\bf
H}^\dagger|\phi_n\rangle=E^*_n{\bf S}|\phi_n\rangle$. Then
\begin{eqnarray}
{\bf \Phi}^{m,\dagger}{\bf H}{\bf \Psi}^n&=&
E_n{\bf \Phi}^{m,\dagger}{\bf S}{\bf \Psi}^n,\label{dum1} \\
{\bf \Psi}^{n,\dagger}{\bf H}^\dagger{\bf \Phi}^m&=& E^*_m{\bf
\Psi}^{n,\dagger}{\bf S}^\dagger{\bf \Phi}^m\label{dum2}
\end{eqnarray}
Taking hermitian conjugate of Eq.(\ref{dum2}),
\begin{eqnarray}
{\bf \Phi}^{m,\dagger}{\bf H}{\bf \Psi}^n&=& E_m{\bf
\Phi}^{m,\dagger}{\bf S}{\bf \Psi}^n \label{dum3}
\end{eqnarray}
From (\ref{dum1}) and (\ref{dum3}), we have
\begin{eqnarray}
{\bf \Phi}^{n,\dagger}{\bf S}{\bf \Psi}^m=C_m\delta_{nm}
\end{eqnarray}
For the normalized wave function $|\psi_n\rangle$ and
$|\phi_n\rangle$,
\begin{eqnarray}
{\bf \Phi}^\dagger{\bf S\Psi}={\bf I}
\end{eqnarray}
It is the usual orthogonality relation for eigenvectors expressed in
a nonorthogonal basis set. For an hermitian Hamiltonian ${\bf
H}={\bf H}^\dagger$, $|\psi_n\rangle=|\phi_n\rangle$, we have
$${\bf \Psi}^\dagger{\bf S\Psi}={\bf I}.$$

\section{Orthogonalize Hamiltonian expressed in nonorthogonal basis}
\label{Orbasis} In this appendix, we will show how to construct a
new orthogonal basis from the atomic real-space nonorthogonal basis.
We will transform the original Hamiltonian ${\bf H}$ which is
expressed in the nonorthogonal basis into Hamiltonian $\tilde{\bf
H}$ expressed in the new orthogonal basis. Of course, instead of
${\bf S}$, the overlap matrix in the new basis will be ${\bf I}$.

Denoting nonorthogonal basis $|\mu\rangle$ and orthogonal basis
$|j\rangle$, they are related by unitary transform operator ${\bf
\mathcal{U}}$
\begin{eqnarray}
&&|{\mu}\rangle=\sum_j|j\rangle\langle j|{\mu}\rangle
=\sum_j|j\rangle{\bf \mathcal{U}}_{j\mu}\nonumber \\
&&{\bf \mathcal{U}}_{j\mu}=\langle j|{\mu}\rangle
\end{eqnarray}
where we have used the completeness of orthogonal basis $|j\rangle$.
Using the orthogonality $\langle i|j\rangle=\delta_{ij}$
\begin{eqnarray}
\sum_{\mu\nu}\langle
i|\mu\rangle\langle\mu|\nu\rangle\langle\nu|j\rangle&=&
\sum_{\mu\nu}{\bf {\mathcal U}}_{i\mu}{\bf S}_{\mu\nu}{\bf {\mathcal
U}}^\dagger_{\nu j}=\delta_{ij}\nonumber
\end{eqnarray}
where we have used the completeness of nonorthogonal basis. In the
matrix form, ${\bf {\mathcal U}}{\bf S}{\bf {\mathcal
U}}^\dagger={\bf I}$. We can formally define $${\bf {\mathcal
U}}={\bf S}^{-\frac{1}{2}},~~{\bf {\mathcal U}}^\dagger=\left[{\bf
S}^{-\frac{1}{2}}\right]^\dagger.$$ Then new Hamiltonian ${\tilde
{\bf H}}$ expressed in basis $|i\rangle$ can be expressed as:
\begin{eqnarray}
{\tilde {\bf H}}_{ij}&=&\langle i|H|j\rangle\nonumber \\
&=&\sum_{\mu\nu}\langle
i|\mu\rangle\langle\mu|H|\nu\rangle\langle\nu|j\rangle\nonumber \\
&=&\sum_{\mu\nu}{\bf {\mathcal U}}_{i\mu}{\bf H}_{\mu\nu}{\bf
{\mathcal U}}^\dagger_{\nu j}
\end{eqnarray}
In the matrix form, ${\tilde {\bf H}}={\bf S}^{-\frac{1}{2}}{\bf
H}\left[{\bf S}^{-\frac{1}{2}}\right]^\dagger$.

We now discuss how to find the matrix ${\bf S}^{-\frac{1}{2}}$.
Without loss generality, we assume the real overlap matrix ${\bf S}$
satisfies eigen function ${\bf SV}={\bf V}{\rm diag}(\lambda_1, ...,
\lambda_n)$ with the eigenvalues $\lambda_1, ..., \lambda_n$ and
eigenvectors ${\bf V}=[v_1, ..., v_n]$. Since ${\bf S}$ is real and
symmetric, the eigenvectors are real and orthogonal, and it thus
holds that ${\bf V}^\dagger {\bf V} = {\bf V} {\bf V}^\dagger = {\bf
I}$. Then
\begin{eqnarray}
{\bf S}&=&{\bf V}{\rm diag}(\lambda_1, ..., \lambda_n){\bf
V}^\dagger\nonumber
\\&=&{\bf V}{\rm diag}(\sqrt{\lambda_1}, ..., \sqrt{\lambda_n}){\bf V}^\dagger
{\bf V}{\rm diag}(\sqrt{\lambda_1}, ..., \sqrt{\lambda_n}){\bf
V}^\dagger\nonumber
\end{eqnarray}
It follows that
\begin{eqnarray}
{\bf S}^{\frac{1}{2}}={\bf V}{\rm diag}(\sqrt{\lambda_1}, ...,
\sqrt{\lambda_n}){\bf V}^\dagger\label{dumA2}
\end{eqnarray}
From ${\bf S}^{-\frac{1}{2}}{\bf S}^{\frac{1}{2}}={\bf I}$ and
Eq.(\ref{dumA2}), we have
\begin{eqnarray}
&&{\bf S}^{-\frac{1}{2}}{\bf V}{\rm diag}(\sqrt{\lambda_1}, ...,
\sqrt{\lambda_n}){\bf V}^\dagger={\bf I}\nonumber \\
&&{\bf S}^{-\frac{1}{2}}{\bf V}{\rm diag}(\sqrt{\lambda_1}, ...,
\sqrt{\lambda_n}){\bf V}^\dagger{\bf V}{\rm
diag}(\frac{1}{\sqrt{\lambda_1}}, ...,
\frac{1}{\sqrt{\lambda_n}}){\bf V}^\dagger\nonumber\\ =&&{\bf
S}^{-\frac{1}{2}}={\bf V}{\rm diag}(\frac{1}{\sqrt{\lambda_1}}, ...,
\frac{1}{\sqrt{\lambda_n}}){\bf V}^\dagger
\end{eqnarray}

In general, the dimension of matrix ${\bf S}$ is infinity, we can't
calculate its eigenvalue $\lambda_i$ and eigenvector $v_i$ by
diagonalizing ${\bf S}$. However, in the tight-binding
representation, the state $\mu$ and $\nu$ hardly overlap when their
separation is large enough in real space, i.e., ${\bf
S}_{\mu\nu}\approx 0$ for most of off-diagonal elements. Considering
the periodic properties in semi-infinite leads, we can select a
block matrix which is large enough to include all the overlap
between leads and central molecular regions. For the non-orthogonal
basis including several unit cell of atomic leads as a buffer layer
into the central scattering region is enough to get a good screening
for dc transport calculation. In transforming the Hamiltonian to the
orthogonal basis needed for ac transport calculation, however, it
turns out that we have to include at least 10 unit cells of atomic
leads into the central scattering region. Partly because the overlap
of orthogonal basis has longer range than that of non-orthogonal
basis. With this large simulation box (finite dimension), we can
calculate the overlap matrix ${\bf S}^{\frac{1}{2}}$ therefore
transform ${\bf H}$ into $\tilde{\bf H}$. The accuracy of
transformed Hamiltonian $\tilde{\bf H}$ should be examined by
comparing dc conductance obtained from the original Hamiltonian
${\bf H}$ and the transformed Hamiltonian $\tilde{\bf H}$.


\end{appendix}
$${\bf ACKNOWLEDGMENTS}$$

This work was supported by a RGC grant (HKU 705409P) from the
government of HKSAR.


\begin{thebibliography}{00}
\bibitem{foot1}
A short account of this method can be found in
Ref.\onlinecite{WangBin}.
\bibitem{foot2}
The exact solution of transient current on the orthogonal basis set
has been derived in Ref.\onlinecite{Maciejko}.
\end{thebibliography}
\end{document}